\documentclass[aps, pra, reprint]{revtex4-2}
\usepackage{graphicx} 
\usepackage{amsmath, physics}
\usepackage{amssymb}
\usepackage{amsfonts}
\usepackage[unicode]{hyperref}
\hypersetup{
   unicode=true,          %
   plainpages=false,
   colorlinks=true,       
   citecolor=blue,        
   urlcolor=blue
}
\usepackage[caption=false,position=top,singlelinecheck=off,justification=raggedright]{subfig}
\usepackage{color}

\begin{document}

\title{Theory of parametric resonance for discrete time crystals in fully connected spin-cavity systems}

\author{Roy D. Jara Jr.}
\email[These authors contributed equally to this work.:]{rjara@nip.upd.edu.ph}
\affiliation{National Institute of Physics, University of the Philippines, Diliman, Quezon City 1101, Philippines}

\author{Dennis F. Salinel}
\email[These authors contributed equally to this work.:]{dsalinel@nip.upd.edu.ph}    
\affiliation{National Institute of Physics, University of the Philippines, Diliman, Quezon City 1101, Philippines}

\author{Jayson G. Cosme}
\affiliation{National Institute of Physics, University of the Philippines, Diliman, Quezon City 1101, Philippines}

\date{\today}

\begin{abstract}
We pinpoint the conditions necessary for discrete time crystal (DTC) formation in fully connected spin-cavity systems from the perspective of parametric resonance by mapping these systems onto oscillator like models. We elucidate the role of nonlinearity and dissipation by mapping the periodically driven open Dicke model onto effective linear and nonlinear oscillator models, while we analyze the effect of global symmetry breaking using the Lipkin-Meshkov-Glick model with tunable anisotropy. We show that the system's nonlinearity restrains the dynamics from becoming unbounded when driven resonantly. On the other hand, dissipation keeps the oscillation amplitude of the period-doubling instability fixed, which is a key feature of DTCs. The presence of global symmetry breaking in the absence of driving is found to be crucial in the parametric resonant activation of period-doubling response. We provide analytic predictions for the resonant frequencies and amplitudes leading to DTC formation for both systems using their respective oscillator models.
\end{abstract}

\maketitle

\section{Introduction}

The emergence and stability of pattern formation is a question of fundamental importance due to its ubiquity in multiple systems across all scales. While spatial order can appear in equilibrium, it can also be induced in resonantly driven systems \cite{chen_amplitude_1999}. Such driving creates parametric instabilities that manifest as an exponential growth of the system's nonzero momentum modes, leading to the formation of crystalline \cite{chen_amplitude_1999, zhang_square_1996} and quasi-crystalline structures \cite{zhang_square_1996}. Spatial order due to parametric instabilities has been observed in both classical fluids \cite{edwards_patterns_1994, torres_five-fold_1995, kudrolli_superlattice_1998} and quantum fluids such as Bose-Einstein condensates \cite{staliunas_faraday_2002, engels_observation_2007, dupont_emergence_2023, wintersperger_parametric_2020, nguyen_parametric_2019, smits_observation_2018}.

Analogous to spatial order, a system can also achieve temporal order by spontaneously breaking its time-translation symmetry. Since its conception \cite{wilczek_quantum_2012}, time crystals have been explored both theoretically and experimentally in multiple platforms, ranging from spin systems \cite{FloquetLMG, Lazarides_2020, II_von, Khemani_PRL, pizzi_higher-order_2021, zhu_dicke_2019, nurwantoro_discrete_2019, munoz-arias_floquet_2022, gong_discrete_2018, krishna_measurement-induced_2023, chan_limit-cycle_2015, owen_quantum_2018, frey_realization_2022, nie_mode_2023, mattes_entangled_2023}, bosonic systems \cite{Pizzi, bakker_driven-dissipative_2022, kesler_observation_2021, tuquero_dissipative_2022, kongkhambut_observation_2022, piazza_self-ordered_2015, kosior_nonequilibrium_2023, johansen_role_2023, gao_self-organized_2023, buca_dissipation_2019, dogra_dissipation-induced_2019}, superconductors \cite{Ojeda_2023, ojeda_collado_emergent_2021, homann_higgs_2020}, and particles bouncing in oscillating mirrors \cite{kuros_phase_2020, giergiel_creating_2020}. Unlike their spatial counterparts, however, this dynamical phase can only occur outside of equilibrium \cite{watanabe_absence_2015}, and thus they require a continuous input of energy to emerge. A system can enter a time crystalline phase by introducing a constant drive \cite{kongkhambut_observation_2022, chen_realization_2023, kesler_emergent_2019, piazza_self-ordered_2015, kosior_nonequilibrium_2023, chan_limit-cycle_2015, buca_dissipation_2019, owen_quantum_2018, johansen_role_2023, gao_self-organized_2023, nie_mode_2023, mattes_entangled_2023, dogra_dissipation-induced_2019} or by periodic driving. The latter scenario results in a discrete time crystal (DTC), which manifests as a subharmonic oscillation of an order parameter. While in general DTCs do not require any additional global symmetry breaking \cite{Floquet, lledo_dissipative_2020}, resonantly driven systems with $\mathbb{Z}_{n}$ symmetry can simultaneously break their time-translation and global symmetry, leading to subharmonic oscillations between the system's static symmetry-broken states \cite{kesler_observation_2021, chitra_dynamical_2015, munoz-arias_floquet_2022, FloquetLMG, Khemani_PRL, Khemani_PRE}. This apparent connection between DTCs and the symmetry-broken states invites the question about the role that global symmetry breaking plays in the emergence of DTCs in certain systems.

Multiple works in various systems suggest an intimate connection and correspondence between DTCs in spin-cavity systems and the subharmonic response in parametric oscillators, similar in spirit to how parametric instabilities lead to spatial order \cite{kongkhambut_realization_2021, skulte_parametrically_2021, kesler_observation_2021, bastidas_entanglement_2010, bastidas_nonequilibrium_2012, chitra_dynamical_2015, cosme_bridging_2023}. A rigorous mathematical correspondence has only been shown in the fully connected spins interacting via photons in a single lossy cavity mode when the Dicke model (DM) is mapped onto a linear oscillator model (LOM) \cite{chitra_dynamical_2015}. However, owing to the linearity of the LOM \cite{kovacic_mathieus_2018}, unbounded dynamics for resonant driving prevent any further analysis of the precise conditions that lead to the emergence of DTCs in driven-dissipative systems. This limitation also restricts what can be known about the role of nonlinearity in the onset of DTCs, which is typically assumed to be present in the mean-field limit of interacting systems hosting DTCs.

\begin{figure}[htbp!]
    \centering
    \includegraphics[scale = 0.34]{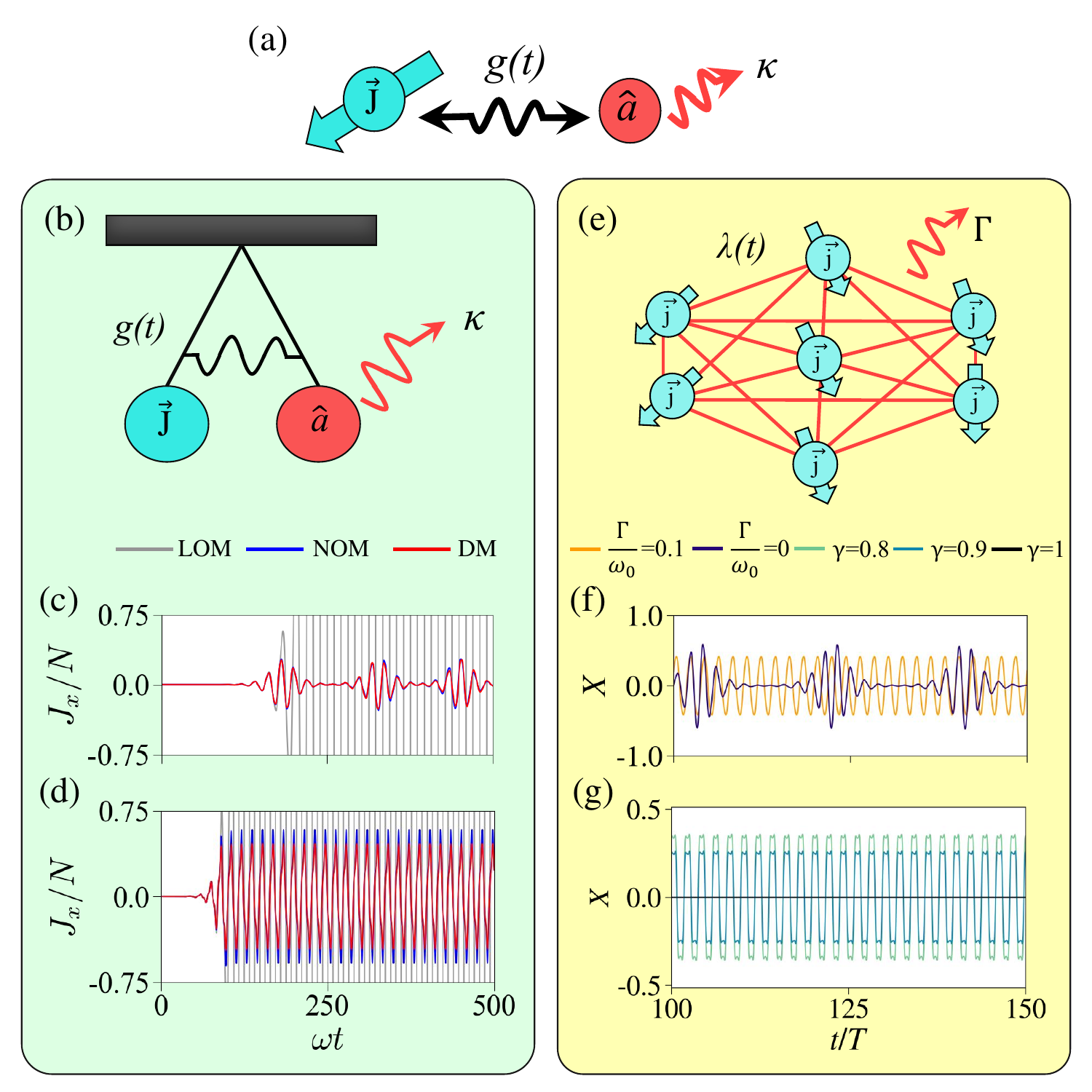}
    \caption{(a) Sketch of a generic fully connected spin-cavity system, where the $N$ two-level systems are represented by a single collective spin. (b) Sketch of the effective parametric oscillator model of a spin-cavity system. Exemplary dynamics of the LOM, NOM, and DM for (c) $\kappa = 0$ and $\{\omega_{d}, A \} = \{0.6 \omega, 0.1  \}$ and (d) $\kappa = 0.5\omega$ and $\{\omega_{d}, A \} = \{0.8 \omega, 0.5 \}$, with $g_{0} = 0.9 g_{c}$. (e) Sketch of the periodically driven open LMG model. Exemplary dynamics of the open LMG are shown for different (f) spin dissipation strengths and (g) anisotropies. The remaining parameters are set to (f) $\{ \lambda_0/\omega_{0}, A, \omega_d/\omega_{0}, \gamma \} = \{ 0.8, 0.2, 1.8, 0 \}$ and (g) $\{ \Gamma/\omega_{0}, \lambda_0/\omega_{0}, A\} = \{ 0.01, 0.8, 0.5\}$ with $\omega_{d} \approx 2\Omega_0$.} 
    \label{fig:main_results}
\end{figure}

Motivated by works on classical time crystals in the coupled nonlinear oscillator model (NOM) \cite{yao_classical_2020, heugel_classical_2019, heugel_role_2023, nicolaou_anharmonic_2021}, we extend the correspondence of DTCs to parametric resonances by mapping a generic spin-cavity system shown in Fig.~\ref{fig:main_results}(a) to an effective NOM [cf. Fig.~\ref{fig:main_results}(b)]. This mapping allows us to elucidate the role of nonlinearity and dissipation on the onset and stability of DTCs. As a test bed, we map the periodically driven DM \cite{emary_chaos_2003, bastidas_nonequilibrium_2012, dimer_proposed_2007} onto both the LOM and NOM and check how nonlinearity and dissipation affect its dynamics. We demonstrate in Figs.~\ref{fig:main_results}(c) and \ref{fig:main_results}(d) that the effective nonlinearity due to the spin-cavity and spin-spin interaction prevents the system's dynamics from becoming unbounded, allowing for more dynamical phases in the parametric instability regions where the LOM breaks down. Meanwhile, the dissipation allows the system to enter a clean period-doubling response in finite time, a key feature of DTCs observed in open DMs \cite{kesler_observation_2021, chitra_dynamical_2015}. Since our focus is on the period-doubling response with constant amplitude of oscillations as predicted in Ref.~\cite{gong_discrete_2018} and then later observed in Ref.~\cite{kesler_observation_2021}, we will use the term DTC to exclusively refer to such a phase in this work.

To explore the role of the global symmetry breaking which is always present in the DM due to its superradiant phase transition \cite{dimer_proposed_2007, emary_chaos_2003}, we also examine the dynamics of a periodically driven open Lipkin-Meshkov-Glick (LMG) model with tunable anisotropy \cite{ParkinsPRA}, as schematically represented in Fig.~\ref{fig:main_results}(e). In this system, the anisotropy in the spin interaction controls the separation between the symmetry-broken states at strong interaction strength \cite{ParkinsPRA}, therefore allowing a systematic study of the role of global symmetry breaking in the onset of DTCs. Unlike previous studies focusing only on the closed LMG for $\gamma = 0$ \cite{cosme_bridging_2023, munoz-arias_floquet_2022, FloquetLMG}, which can be obtained from the DM after adiabatically eliminating the photonic mode in the bad cavity limit, where the cavity dissipation strength approaches infinity, we include not only a tunable anisotropy but also dissipation through global spin decay. We show in Fig.~\ref{fig:main_results}(f) that spin dissipation allows for the DTC formation in the LMG model, similar to the open DM. On the other hand, we observe that the DTC response amplitude vanishes in the isotropic limit $\gamma = 1$, coinciding with the limit where the system does not break its $\mathbb{Z}_{2}$ symmetry even for large spin coupling strength $\lambda$, as shown in Fig.~\ref{fig:main_results}(g). Thus, our results establish the conditions needed to form stable DTCs via parametric resonance in fully connected spin-cavity systems.

The paper is structured as follows. In Sec. \ref{sec:open_dm} we introduce the open DM and its mapping to effective oscillator models. We demonstrate here how the interplay of nonlinearity and dissipation modifies the phase diagram of the DM's effective oscillator models, and how these lead to stable DTCs. In Sec. \ref{sec:open_lmg} we introduce the open LMG model and derive its steady states for different anisotropy values. We also show how the anisotropy is linked with $\mathbb{Z}_{2}$ symmetry breaking and present its role in the DTC formation in the LMG model. We provide a summary and possible extensions of our work in Sec. \ref{sec:conclusion}.

\section{Open Dicke Model}\label{sec:open_dm}

We first consider the open DM described by the Lindblad master equation \cite{dimer_proposed_2007} 
\begin{equation}
    \label{eq:dicke_master_eq}
    \partial_{t} \hat{\rho} = -i \left[ \frac{\hat{H}_{\mathrm{DM}}}{\hbar} , \rho \right] + \kappa \left(2\hat{a} \hat{\rho} \hat{a}^{\dagger} - \left\{ \hat{a}^{\dagger}\hat{a}, \hat{\rho}  \right\} \right),
\end{equation}
where
\begin{equation}
    \label{eq:dicke_model}
    \frac{\hat{H}_{\mathrm{DM}}}{\hbar} = \omega \hat{a}^{\dagger}\hat{a} + \omega_{0}\hat{J}_{z} + \frac{2g}{\sqrt{N}} \left(\hat{a} + \hat{a}^{\dagger} \right) \hat{J}_{x} .
\end{equation}
The open DM describes the interaction between $N$ two-level systems, represented by the collective spin operators $\hat{J}_{x, y, z}$ and a single lossy cavity with dissipation strength $\kappa$, represented by the bosonic annihilation (creation) operator $\hat{a}$ ($\hat{a}^{\dagger}$) \cite{dimer_proposed_2007}. The photonic and atomic transition frequencies are $\omega$ and $\omega_{0}$, respectively, while $g$ is the light-matter coupling.

In equilibrium, the open DM has two phases: the normal phase (NP) and the superradiant (SR) phase \cite{dimer_proposed_2007}. The NP is characterized by a fully polarized collective spin in the $-z$ direction and a cavity in a vacuum state. The SR phase, on the other hand, is associated with the global $\mathbb{Z}_{2}$ symmetry breaking of the system manifesting in the nonzero photon number in the cavity and finite-$x$ component in the collective spins. The critical point \cite{dimer_proposed_2007} 
\begin{equation}
    \label{eq:dicke_crit_point}
    g_{c} = \frac{1}{2} \sqrt{\frac{\omega_{0}}{\omega} \left( \kappa^{2} + \omega^{2} \right)  }
\end{equation}
separates these two phases. The open DM also exhibits a DTC phase when the light-matter coupling is periodically driven \cite{chitra_dynamical_2015, kesler_observation_2021, cosme_bridging_2023}. Here we consider a periodic driving of the form
\begin{equation}
    \label{eq:dicke_driving}
    g(t) = g_{0}\left[1 + A \sin(\omega_{d}t) \right],
\end{equation}
where $g_{0}$ is the static light-matter coupling, $A$ is the driving amplitude, and $\omega_{d}$ is the driving frequency. In the following, we consider $\omega = \omega_{0}$ and $g_{0} = 0.9g_{c}$.

Following the mean-field approach, which is exact for the DM in the thermodynamic limit $N \rightarrow \infty$, \cite{carollo_exactness_2021}, we assume that quantum fluctuations are negligible so that $\left< \hat{A}\hat{B} \right> \approx \left< \hat{A} \right>\left< \hat{B} \right>$. This allows us to treat the cavity mode as a complex number $a = \left< \hat{a} \right>$ and the collective spin components as real numbers $J_{x, y, z} = \left< \hat{J}_{x, y, z} \right>$. Finally, the system will be initialized close to the steady state of the NP \cite{dimer_proposed_2007},
\begin{subequations}
\label{eq:dicke_initial_states}
    \begin{equation}
        \label{eq:cavity_initial_states}
        a = \epsilon
    \end{equation}
    \begin{equation}
        \label{eq:Jx_Jy_Jz_initial_state}
        J_{x} = - \epsilon \sqrt{N}, \quad J_{y} = 0 , \quad J_{z} = -\frac{N}{2} \sqrt{1 -  \frac{4 \epsilon^{2} }{N}  },
    \end{equation}    
\end{subequations}
where $\epsilon$ is a perturbation set to $\epsilon = 0.01$. In a moment, we will show that this is similar to setting the initial states of the effective oscillator model at its stable fixed point when $A=0$.

\subsection{Effective Oscillator Models}

To reduce the open DM into an effective oscillator model (OM), we use the Holstein-Primakoff representation \cite{emary_chaos_2003, vogl_resummation_2020, carollo_non-gaussian_2023, boneberg_quantum_2022} 
\begin{equation}
    \label{eq:hp_representation}
    \hat{J}_{z} =  \hat{b}^{\dagger}\hat{b} - \frac{N}{2}, \quad \hat{J}_{-} = \hat{J}_{+}^{\dagger} = \sqrt{N}\sqrt{1 - \frac{\hat{b}^{\dagger}\hat{b}}{N}}\hat{b}
\end{equation}
to convert the collective spin operators into a bosonic operator associated with the atomic excitation of the system, and take the thermodynamic limit. In this limit, we can expand the operator $\hat{F} = \sqrt{1 - \hat{b}^{\dagger} \hat{b} / N}$, using its Taylor series expansion
\begin{equation}
    \label{eq:hpr_expansion}
    \sqrt{1 - \frac{\hat{b}^{\dagger}\hat{b}}{N}} \approx 1 - \frac{1}{2}\frac{\hat{b}^{\dagger}b}{N} - \frac{1}{8}\left( \frac{\hat{b}^{\dagger}b}{N} \right)^{2} - \ldots,
\end{equation}
and truncate it up to an arbitrary term depending on the accuracy needed. For our case, we consider both the zeroth-order truncation $\hat{F}^{(0)} = 1$ and the first-order truncation $\hat{F}^{(1)} = 1 - \hat{b}^{\dagger} \hat{b} / 2N$ of $\hat{F}$. Substituting $\hat{F}^{(0)}$ in Eqs.~\eqref{eq:hp_representation} and \eqref{eq:dicke_model}, we retrieve the effective LOM of the closed DM \cite{emary_chaos_2003},
\begin{equation}
    \label{eq:dicke_lom_hamiltonian}
    \frac{\hat{H}_{\mathrm{LOM}}}{\hbar} = \omega \hat{a}^{\dagger}\hat{a} + \omega_{0} \hat{b}^{\dagger}\hat{b} + g(t) \left( \hat{a}^{\dagger} + \hat{a} \right) \left( \hat{b} + \hat{b}^{\dagger} \right).
\end{equation}
Meanwhile, substituting $\hat{F}^{(1)}$ leads to NOM of the form
\begin{equation}
    \label{eq:dicke_nom_hamiltonian}
    \frac{\hat{H}_{\mathrm{NOM}}}{\hbar} = \frac{\hat{H}_{\mathrm{LOM}}}{\hbar} - \frac{g(t)}{2N} \left( \hat{a} + \hat{a}^{\dagger} \right) \left( \hat{b}^{\dagger}\hat{b}\hat{b} + \hat{b}^{\dagger}\hat{b}^{\dagger}\hat{b} \right),
\end{equation}
where the nonlinearity appears as a perturbation in the original LOM Hamiltonian. Note that the approximation for the two models is only valid in the NP, where both the cavity and atomic modes do not have any macroscopic occupations \cite{emary_chaos_2003}. Given this constraint, we restrict our attention to $g_{0} < g_{c}$ so that the system remains in the NP at the initial time $t_{i}=0$. We also consider the mean-field dynamics of both the LOM and the NOM such that we can treat both atom and cavity modes as complex numbers, i.e., $a \equiv \left< \hat{a} \right>$ and $b \equiv \left< \hat{b} \right>$ with $a, b \in \mathbb{C}$.

We initialize the two OMs near their stable fixed point in the undriven case $a_{0} = -b_{0} = \epsilon$. In terms of the collective spin components, this corresponds to setting the initial states to Eq.~\eqref{eq:Jx_Jy_Jz_initial_state} since in the thermodynamic limit $J_{x} \approx  \sqrt{N} \Re(b)$ and $J_{y} \approx \sqrt{N} \Im(b)$, assuming $\hat{F} \approx \hat{F}^{(0)}$. Note that so long as $\epsilon$ is small, the system should remain close to the steady state of the NP at $t_{i} = 0 \omega^{-1}$. We then numerically integrate the equations of motions of the LOM and the NOM, given by Eqs.~\eqref{eq:lom_atom_cavity_eom} and \eqref{eq:nom_atom_cavity_eom}, respectively, in Appendix \ref{sec:om_eom}, using the fourth-order Runge-Kutta algorithm. Note that we will use the same integration scheme to probe the dynamics of the LMG model in the latter part of this paper. To capture the experimentally accessible long-time behavior of the system, we set the final time of the simulation to $t_{f} = 1000 \omega^{-1}$. We present in Figs.~\ref{fig:main_results}(c) and \ref{fig:main_results}(d) the exemplary resonant dynamics of the $J_{x}$ of the LOM, NOM, and DM for $\kappa = 0$ and $\kappa = 0.5 \omega$, respectively. We can see that for both values of $\kappa$, the dynamics of the LOM can be characterized by a period-doubled oscillation with an exponentially growing amplitude. As we introduce nonlinearity in the OM, the dynamics of the oscillators become bounded as seen in the behavior of $J_{x}$. In particular, for $\kappa = 0$, we observe a beating oscillation in both the NOM and DM. On the other hand, the response amplitude approaches a constant value in the long-time limit when $\kappa$ becomes nonzero. These results highlight the capability of the NOM to predict the qualitative behavior of the dynamics of the DM initialized in the NP. In particular, the mere inclusion of the first nonlinear term of $\hat{F}$ already provides a good approximation of the DM's dynamics. This result remains true for all values of driving amplitude considered in the main text, i.e., from $A = 0$ to $A = 1$, as we demonstrate in Appendix~\ref{sec:appendix_breakdown}. Finally, note that for the given parameters in Fig.~\ref{fig:main_results}(d), the dynamics of $J_{x}$ for both the NOM and DM is a period-doubling instability; hence we classify them as DTCs.

To map out the values of $\omega_{d}$ and $A$ where the OM enters the DTC phase, we construct the $\omega_{d}$-$A$ phase diagram of both the LOM and the NOM. We first classify the dynamics of the oscillator models as bounded or unbounded (UB) by calculating the maximum response amplitude $\max\{J_{x}\} / N$. If $\max \{ J_{x} \} / N \leq 10^{-3} $, the system is in the NP, while the dynamics is UB if it exceeds the physically allowed maximum value of a spin component $\max \{ J_{x} \} / N > 1$. For dynamics with intermediate values of $\max \{ J_{x} \} / N$, we further distinguish the DTC phase corresponding to a constant-amplitude subharmonic oscillation from other nontrivial dynamics by first calculating the time-averaged decorrelator \cite{pizzi_higher-order_2021, Pizzi}
\begin{equation}
    \label{eq:decorrelator}
    d^{2} = \frac{1}{N_{\mathrm{step}}} \sum_{t = t_{i}}^{t_{f}} \sum_{i}^{n} \left[ \left|\mathcal{O}_{i}^{\mathrm{o}}(t) \right|^{2} - \left| \mathcal{O}_{i}^{\mathrm{p}}(t) \right|^{2}  \right]^{2},
\end{equation}
where $N_{\mathrm{step}}$ is the number of steps from the initial to the final time of the simulation. The order parameters $\mathcal{O}^{\mathrm{o}}_{i}$ and $\mathcal{O}^{\mathrm{p}}_{i}$ track the system's original and perturbed dynamics initialized near $a_{0}$ and $b_{0}$, respectively. Equation \eqref{eq:decorrelator} measures the deviation of two initially close dynamics as $t \rightarrow \infty$ \cite{Pizzi, pizzi_higher-order_2021}. A value of $d^{2} \ll 1$ implies that the two dynamics converge in the long-time limit and thus the system's dynamics is nonchaotic. Here we consider $\mathcal{O}_{1}^{\mathrm{O, P}} = J_{x} / N$ and $\mathcal{O}_{2}^{\mathrm{O, P}} = 2 \Re(a) / \sqrt{N}$ as our order parameters and initialize the perturbed dynamics to $a^{\mathrm{p}}_{0} = a_{0} + \delta$ and $b^{\mathrm{p}}_{0} = b_{0} + \delta$, with $\delta = 10^{-6}$. We then identify the clean oscillations with constant amplitude by calculating the standard deviation of the amplitude envelope of $J_{x} / N$, $\sigma_{\mathrm{amp}}$. We also take the Fourier transform of $J_{x}$ within the time window $t = t_{f} - 10T$ to $t = t_{f}$, with $T$ the driving period. The system's dynamics is a DTC if $\sigma_{\mathrm{amp}} \ll 1$, and the response frequency associated with the highest peak of its frequency spectrum is $\omega_{d} / n$, where $n \geq 2$ is integer valued.

\begin{figure}[t]
    \includegraphics[scale = 0.4]{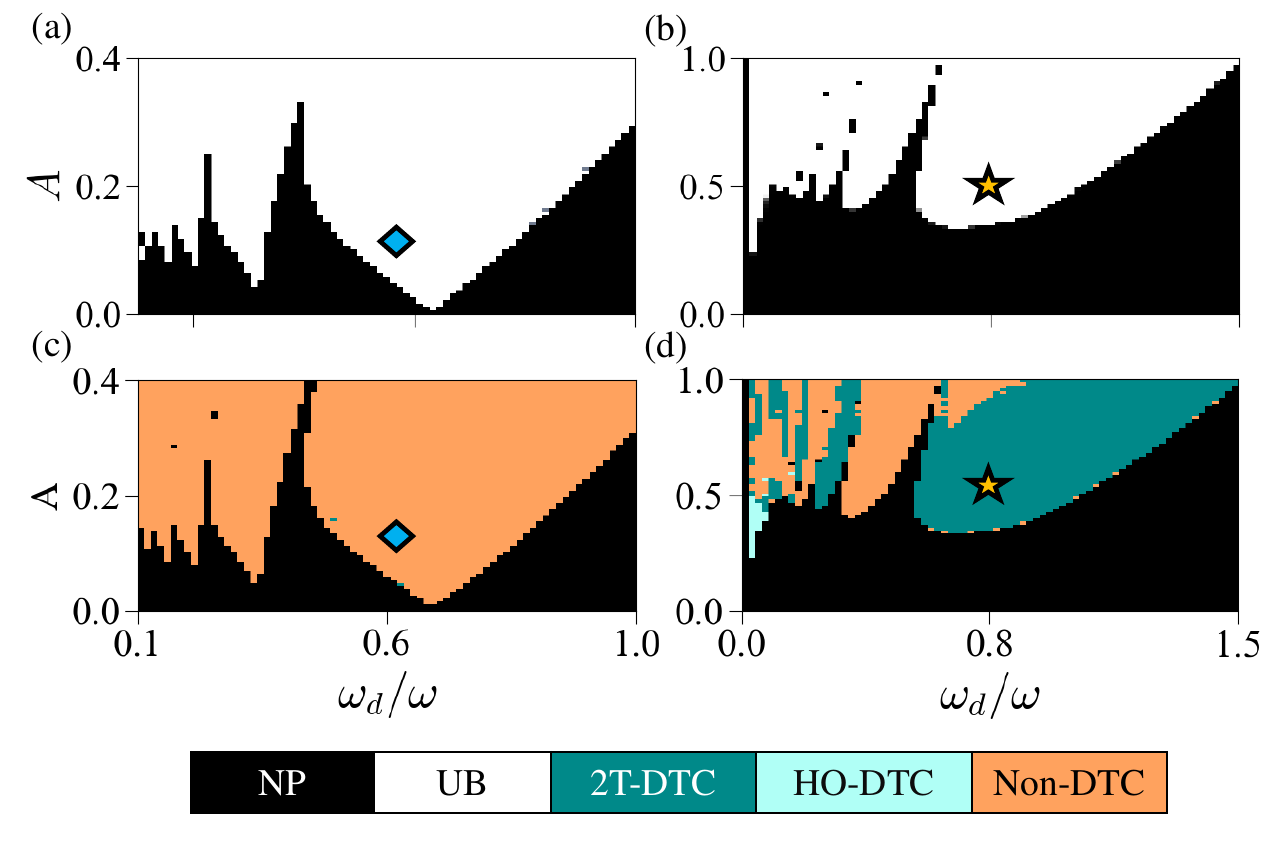}
    \caption{The $\omega_{d}$-$A$ phase diagram of the LOM for (a) $\kappa = 0$ and (b) $\kappa = 0.5 \omega$. Phase diagram of the NOM for (c) $\kappa = 0$ and (d) $\kappa = 0.5 \omega$. The diamond (star) corresponds to the exemplary dynamics of the closed (open) OMs shown in Fig.~\ref{fig:main_results}(c) [Fig.~\ref{fig:main_results}(d)]. The remaining parameter is set to $g_{0} = 0.9 g_{c}$.}
    \label{fig: oscillator phase diagram}
\end{figure}

We present in Figs.~\ref{fig: oscillator phase diagram}(a) and \ref{fig: oscillator phase diagram}(b) the LOM phase diagram for $\kappa = 0$ and $\kappa = 0.5 \omega$, respectively. For both values of $\kappa$, the LOM has only two phases: the NP and the UB dynamics lying in the resonance lobes of the system. The observed UB dynamics are period doubled and have an exponentially growing amplitude as shown in Figs.~\ref{fig:main_results}(c) and \ref{fig:main_results}(d). This is consistent with the results for parametrically driven linear oscillators \cite{kovacic_mathieus_2018} and the effective LOM of the Dicke model \cite{chitra_dynamical_2015, bastidas_nonequilibrium_2012}. Meanwhile, we demonstrate in Figs.~\ref{fig: oscillator phase diagram}(c) and \ref{fig: oscillator phase diagram}(d) that by the simple inclusion of the first nonlinear term in the expansion of $\hat{F}$, we get a richer phase diagram for the NOM for $\kappa = 0$ and $\kappa = 0.5 \omega$, respectively. In particular, for $\kappa = 0$, the regions containing UB dynamics are replaced by bounded dynamical phases, all of which are non-DTC phases. As we introduce dissipation, the $2T$-DTC starts to appear in the NOM phase diagram, with the dynamical phase being more prominent in the largest resonant lobe of the system. As shown in Fig.~\ref{fig:main_results}(d), the $2T$-DTC phase manifests as a constant-amplitude period-doubling response that is stable in the long-time limit. We also observe higher-order (HO) DTCs at small values of $\omega_{d}$ and $A$. They correspond to subharmonic oscillations with a response period greater than $2T$.

Our results in  Figs. \ref{fig:main_results}(c), \ref{fig:main_results}(d), \ref{fig: oscillator phase diagram}(c) and \ref{fig: oscillator phase diagram}(d) highlight two of the main conclusions of this work: For DTCs to form in a fully connected spin-cavity system in the mean-field level, the system must have an effective nonlinearity coming from either the spin-cavity or spin-spin interaction, and a dissipation channel. The effective nonlinearity due to the interaction prevents the system from becoming unbounded when driven at parametric resonance, while the dissipation allows the amplitude of the system's dynamics to approach a constant value at a finite time. Without both, resonant parametric driving would only lead to either unbounded, and therefore unphysical, dynamics or a non-DTC phase.

\subsection{Parametric Resonances of the OM}\label{sec:parametric_resonance}

\begin{figure*}
    \centering
    \includegraphics[scale = 0.36]{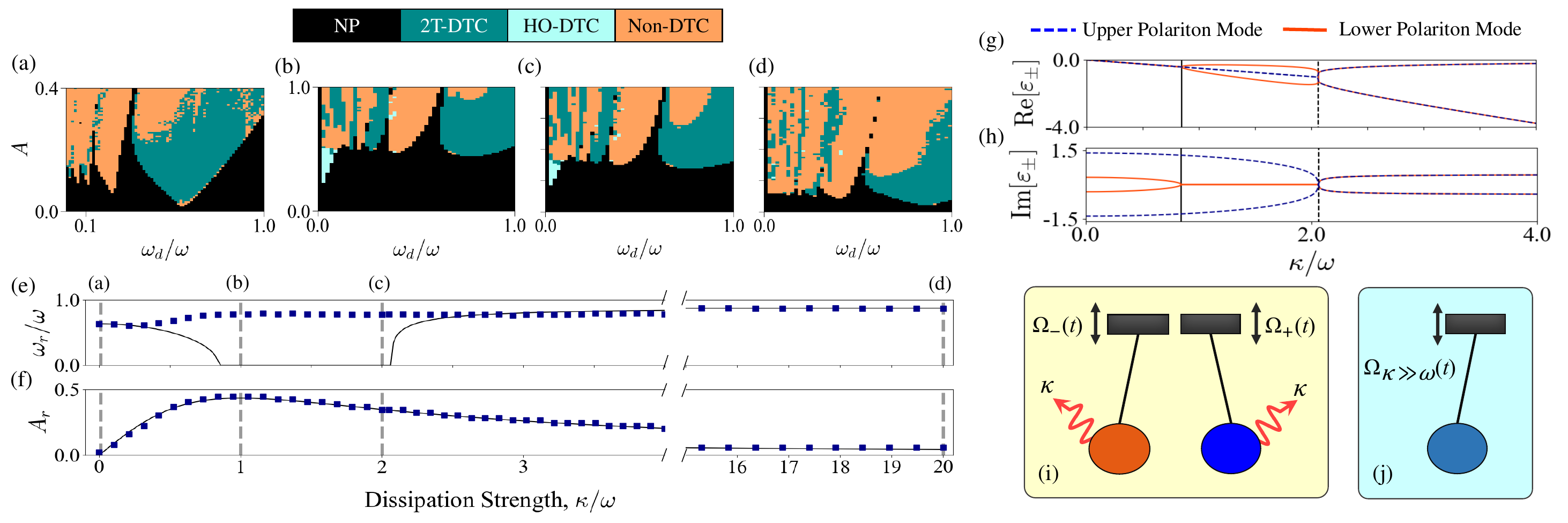}
    \caption{(a)--(d) Phase diagram of the NOM for increasing values of dissipation strength. The values of $\kappa$ considered are $\kappa = \{0.1 \omega, 1.0 \omega, 2.0 \omega, 20 \omega\}$. Trends of the simulated (e) resonant frequency and (f) minimum resonant amplitude are plotted as a function of $\kappa$. In (e) the solid lines correspond to the lower polariton frequency $2\Omega_{-}$, while in (f) they correspond to the analytic expression of $A_{r}$ derived in Ref. \cite{jager_dissipative_2023}. The phase diagrams in (a)--(d) are identified in (e) and (f) using the gray dashed lines. (g) Real and (h) imaginary components of the eigenmodes of the open LOM as a function of $\kappa$. The black solid line corresponds to $\kappa'_{c}$, while the dashed line represents $\kappa_{c}''$. (i)-(j) The effective oscillator picture of the polariton modes is sketched for (i) $\kappa < \kappa_{c}'$ and (j) $\kappa \gg \kappa_{c}''$. The remaining parameter is set to $g_{0} = 0.9 g_{c}$.}
    \label{fig: resonant frequency trend}
\end{figure*}

Given that the DTCs only emerge when dissipation is present in the system, it is natural to ask how varying $\kappa$ alters the phase diagram of the NOM. We show in Figs.~\ref{fig: resonant frequency trend}(a)--\ref{fig: resonant frequency trend}(d) the $\omega_{d}$-$A$ phase diagram of the NOM for different values of $\kappa$. We can see that the largest resonance lobe shifts towards increasing $\omega_{d}$ as $\kappa$ increases. It then saturates at a constant value at larger $\kappa$, as further highlighted in Fig.~\ref{fig: resonant frequency trend}(e), where we plot the resonant frequency $\omega_{r}$ corresponding to the $\omega_{d}$ where the tip of the largest lobe lies.  We also observe that the minimum resonant amplitude $A_{r}$ of the largest lobe increases with $\kappa$, leading to the shrinking of DTC regions in the phase diagram for our given range of $\omega_{d}$ and $A$. This behavior implies that for finite dissipation strength, the system would require a stronger drive to enter the DTC phase even when it is driven at $\omega_{r}$, consistent with the expected behavior of dissipative parametric oscillators \cite{kovacic_mathieus_2018}. Notice, however, that this remains true only until $\kappa = 1.0\omega$, where $A_{r}$ is at its maximum, as further highlighted in Fig.~\ref{fig: resonant frequency trend}(f). Beyond that, the largest lobe starts to approach $A = 0$ again as $\kappa \rightarrow \infty$, increasing its area as well.

To understand the behavior of $\omega_{r}$ and $A_{r}$, we analytically derive the $\omega_{r}$ of the open NOM by calculating its eigenfrequencies, $\Omega$, and observe how it changes as we vary $\kappa$. The resonant frequency associated with the largest resonance lobe, which we now refer to as the primary resonance, is related to $\Omega$ through the simple relation $\omega_{r} = 2\Omega$ \cite{kovacic_mathieus_2018}. Without loss of generality, we will assume that the LOM and the NOM have the same $\omega_{r}$ and $A_{r}$. This assumption is supported by the phase diagrams in Fig.~\ref{fig: oscillator phase diagram}, where the shape of the resonant lobes for the LOM are identical to those of the NOM for both $\kappa$ considered. As such, let us now consider the equations of motion for the LOM in terms of the pseudo position variables $x$ and $y$ (see Appendix~\ref{sec:om_eom} for details), 
\begin{subequations}
    \label{eq:lom_in_xy}
    \begin{equation}
        \ddot{x} + 2 \kappa \dot{x} = - \left( \omega^{2} + \kappa^{2} \right) x - 2 g \omega y, 
    \end{equation}
    \begin{equation}
        \ddot{y} = - 2 g \omega x - \omega^{2} y.
    \end{equation}
\end{subequations}
Under the series of transformations detailed in Appendix~\ref{subsec:diagonalization}, the equations of motion for the LOM can be diagonalized into two uncoupled oscillators of the form
\begin{equation}
    \label{eq:driven_lom_diagnoalized}
    \ddot{X} + \Omega_{\pm}^{2}(t) X = 0,
\end{equation}
where
\begin{equation}
    \label{eq:driven_polariton_mode}
    \begin{aligned}
    \Omega_{\pm}^{2}(t) &=  \omega^{2} - \frac{\kappa^{2}}{4} \\ 
    & \pm  \omega \sqrt{g'^{2}\left( \omega^{2} + \kappa^{2} \right) - \kappa^{2}} \sqrt{1 + \delta A\sin \left( \omega_{d}t \right)} 
    \end{aligned}
\end{equation}
is the periodically driven polariton frequencies, $g' = g_{0} / g_{c}$, and
\begin{equation}
    \label{eq:drive_amp_scaling}
    \delta = \frac{g'^{2} \left( \kappa^{2} + \omega^{2} \right)}{ g'^{2} \left(\kappa^{2} + \omega^{2} \right) - \kappa^{2}}.
\end{equation}

Restricting our attention to the lower polariton frequency $\Omega_{-}(t)$, we assume that $A \ll 1$, so $\Omega_{-}(t)$ can be approximated as
\begin{equation}
    \label{eq:approx_driven_polariton_mode}
    \Omega^{2}_{-}(t) \approx \Omega_{-}^{2} \pm 2 \delta A \sin \left( \omega_{d}t \right),
\end{equation}
where
\begin{equation}
    \label{eq:lower_polariton_modes}
    \Omega_{-} = \left( \omega^{2} - \frac{\kappa^{2}}{4} -  \omega \sqrt{g'^{2}(\omega^{2} + \kappa^{2}) - \kappa^{2}    }  \right)^{1 / 2}
\end{equation}
is the static lower polariton frequency. Immediately, we see that $\omega_{r} = 2\Re(\Omega_{-})$. As we demonstrate in Fig.~\ref{fig: resonant frequency trend}(e), the analytic $\omega_{r}$ is in good agreement with the numerical $\omega_{r}$ for the limits $\kappa \rightarrow 0$ and $\kappa \rightarrow \infty$. For finite values of $\kappa$, we observe a large deviation between the numerical and analytical $\omega_{r}$, with the analytical $\omega_{r}$ dropping to zero.

To see where this deviation comes from, we plot in Figs.~\ref{fig: resonant frequency trend}(g) and \ref{fig: resonant frequency trend}(h) the real and imaginary components of the complete eigenmodes of the open LOM, $\varepsilon_{\pm}$, respectively, as a function of $\kappa$. The real component of $\varepsilon_{-}$ ($\varepsilon_{+}$) corresponds to the effective dissipation experienced by the lower (upper) polariton mode, while $\Im(\varepsilon_{\pm}) = \Re(\Omega_{\pm})$. Notice that within the interval $\kappa'_{c} \leq \kappa \leq \kappa_{c}''$, where 
\begin{equation}
	\label{eq:first_kappa_crit}
	\kappa_{c}' = 2 \omega \left( 2 g'^{2} - 1 - g' \sqrt{4g'^{2} - 3}  \right)^{1/2}  
\end{equation}
and
\begin{equation}
	\label{eq:second_kappa_crit}
	\kappa_{c}'' = \frac{\omega}{ \sqrt{ 1 / g'^{2} - 1 } }
\end{equation}
are the critical values of $\kappa$ highlighted by the solid and dashed lines, respectively, in Fig.~\ref{fig: resonant frequency trend}(g) and \ref{fig: resonant frequency trend}(h), $\Im(\varepsilon_{-})$ drops to zero, consistent with our result in Fig.~\ref{fig: resonant frequency trend}(e). The contribution of the lower polariton frequency is then converted to an effective dissipation, turning the NOM into an overdamped oscillator. As a result, the approximation done in Eq.~\eqref{eq:approx_driven_polariton_mode} breaks down and the lower polariton mode becomes effectively driven by a nonlinear drive. This explains the sudden increase of the simulated $\omega_{r}$ after passing $\kappa_{c}'$ and its saturation at a constant value while the analytical $\omega_{r}$ is zero. Finally, as $\kappa > \kappa_{c}''$, $\Re(\Omega_{-})$ becomes nonzero again and begins to approach the frequency
\begin{equation}
    \label{eq:polariton_large_kappa}
    \Re(\Omega_{-}) = \Omega_{\kappa \gg \omega}= \omega \sqrt{1 - g'^{2}},  \quad \kappa \gg \omega,
\end{equation}
which is consistent with the behavior of the numerical $\omega_{r}$ for large $\kappa$. Note that similar results were found in the dissipative atom-only description of the DM derived in Ref. \cite{jager_dissipative_2023}.

Now, unlike $\omega_{r}$, finding an analytic expression for $A_{r}$ from Eq.~\eqref{eq:lom_in_xy} is nontrivial since the multiscale analysis motivating the approximation done in $\Omega_{-}(t)$ assumes that $\kappa \rightarrow 0$ \cite{kovacic_mathieus_2018}. As such, the perturbative method cannot access the behavior of the LOM in the large-$\kappa$ limit. Note that this problem can be generally resolved using the classical Floquet theory, where both $\omega_{r}$ and $A_{r}$ can be obtained from the eigenvalues of the LOM's infinite-dimensional Hill's matrix \cite{kovacic_mathieus_2018, nicolaou_anharmonic_2021}. In Ref.~\cite{jager_dissipative_2023} an alternative method for solving $A_{r}$ was proposed by constructing an effective dissipative spin model for the open DM, leading to an OM that can be driven resonantly at a frequency $\omega_{d, R} = 2 \Omega_{\kappa \gg \omega}$. At the same time, its resonant amplitude is given by 
\begin{equation}
    \label{eq:analytic_resonant_amplitude}
    A_{r} = 2 \Omega_{\kappa \gg \omega} \frac{\kappa}{ \omega^{2} + \kappa^{2}}.
\end{equation}
In Fig.~\ref{fig: resonant frequency trend}(f) we plot the numerical $A_{r}$ of the NOM and Eq.~\eqref{eq:analytic_resonant_amplitude} and find excellent agreement between the two. This result implies that there is a value of $\kappa$ where we get the least amount of DTC in the phase diagram due to the large $A_{r}$ needed to enter the DTC phase. We infer this value of $\kappa$ by maximizing Eq.~\eqref{eq:analytic_resonant_amplitude}, giving us $\kappa_{\max} = 1.0\omega$. Beyond $\kappa_{\max}$, both the LOM and the NOM will have low $A_{r}$ and thus a larger region in the phase diagram with DTCs.

We finally conclude our analysis of the open Dicke model by noting that as the system approaches the bad cavity limit, the area of the non-DTC regions in the phase diagram steadily increases together with $\kappa$. This behavior, together with $A_{r} \rightarrow 0$ as $\kappa \rightarrow \infty$, is similar to what we observe when we approach the closed limit $\kappa \rightarrow 0$. We can understand this by considering the effective oscillator picture of the polariton modes for different dissipation strengths, which we infer from the behavior of $\Re(\varepsilon_{\pm})$ and $\Im(\varepsilon_{\pm})$ shown in Fig.~\ref{fig: resonant frequency trend}(g) and \ref{fig: resonant frequency trend}(h), respectively. For weak dissipation strength $\kappa < \kappa'_{c}$, the polariton modes are identified by their distinct frequencies and experience the same linear dissipation strength. As such, we can model them as two uncoupled parametric oscillators with weak damping, as illustrated in Fig.~\ref{fig: resonant frequency trend}(i). On the other hand, the polariton frequencies are degenerate for large dissipation strength $\kappa > \kappa_{c}''$. Additionally, the upper branches of the eigenmodes, $\varepsilon_{\pm}^{\mathrm{U}}$, become nondissipative, with $\Re(\varepsilon_{\pm}^{\mathrm{U}}) \rightarrow 0$, while the lower branches return to a linear damping. As a result, only the upper branches become excited when the open NOM is driven resonantly, while the large dissipation suppresses the lower branches. Thus, the NOM can be treated as a single oscillator with negligible dissipation in the bad cavity limit, which we schematically represent in Fig.~\ref{fig: resonant frequency trend}(j).

\section{Open Lipkin-Meshkov-Glick Model}\label{sec:open_lmg}

Using the LMG model, we now explore the connection between the presence of global symmetry breaking in the undriven limit of the system and emergence of a time crystalline response via parametric resonance. Recall that adiabatic elimination of the cavity mode in Eq.~\eqref{eq:dicke_model} leads to the effective Dicke Hamiltonian in the bad cavity limit as \cite{ParkinsPRA}
\begin{equation}
    \frac{\hat{H}_{\kappa \to \infty}}{\hbar} = \omega_0 \hat{J}_z - \frac{4g^2}{N \omega} \hat{J}_x^2.
\end{equation}
This closely resembles the anisotropic case of the general LMG model, defined by the Hamiltonian \cite{ParkinsPRA, ParkinsPRL, Zhou}
\begin{equation}\label{eq:LMG}
    \frac{\hat{H}_{\text{LMG}}}{\hbar} = \omega_0 \hat{J}_z - \frac{\lambda}{N} \bigl( \hat{J}_x^2 + \gamma \hat{J}_y^2 \bigr), 
\end{equation}
which describes a set of $N$ spin-$1/2$ particles experiencing infinite-range interaction with strength $\lambda$, with $\gamma \in [ -1, 1 ]$ a dimensionless anisotropy parameter. Unlike the DM, we show that the global $\mathbb{Z}_{2}$ symmetry breaking in the general LMG model can be turned off by tuning the anisotropy parameter. This then allows us to systematically study the role of global symmetry in the onset and stability of DTCs in fully connected spin-cavity systems.

In this work we follow the formalism used in Ref.~\cite{ParkinsPRA} and consider an open LMG model in the context of a ring-cavity-QED setup. Such a system can be described by a generalized Lindblad master equation of the form
\begin{equation}
    \partial_t \hat{\rho} = -i \biggl[ \frac{\hat{H}_{\text{LMG}}}{\hbar}, \rho \biggr] + \sum_k \frac{\Gamma_k}{N} \bigl( \hat{X}_k \hat{\rho} \hat{X}_k^{\dagger} - \frac{1}{2} \{ \hat{X}_k^{\dagger} \hat{X}_k, \hat{\rho} \} \bigr),
\end{equation}
where $\Gamma_k$ is the dissipation rate due to channel $k$, and $\hat{X}_k = \alpha_k \hat{J}_+ + \beta_k \hat{J}_-$ is an operator that depends on $\gamma$ ($\alpha_k, \beta_k \in [-1, \, 1]$). This dependence raises the need to specify $\hat{X}_k$ for different anisotropy parameter values and therefore requires multiple master equations to describe different $\gamma$ cases \cite{ParkinsPRA, Zhou}. The abovem-entioned restriction has limited the exploration of the $\mathbb{Z}_2$-symmetry-breaking behavior to the $\gamma = 0$ case in previous studies.

In the mean-field level, two phases have been shown to exist in the LMG model for $\gamma = 0$: the normal phase, which corresponds to the collective spin relaxing towards the lowest-energy configuration $\langle \hat{J}_x \rangle = 0$, and the symmetry-broken (SB) phase, where the collective spin relaxes away from the lowest-energy state $\langle \hat{J}_x \rangle \neq 0$. The latter is a counterpart to the SR phase of the DM in that both phases constitute a broken $\mathbb{Z}_2$ symmetry in the system. As mentioned earlier, the DM reduces to the $\gamma = 0$ LMG model in the bad cavity limit, pointing to the connection between the SR and SB phases. Note that the signs of system parameters were set such that the spin-down configuration is the lowest-energy state of the system. As such, one can alternatively describe the NP as having the spins fully polarized in the $-z$ direction ($\langle \hat{J}_z \rangle = -N/2$) and the SB phase as having the spins polarized at an angle relative to the spin-down configuration ($\langle \hat{J}_z \rangle \neq -N/2$), analogous to the definitions of the NP and SR phases of the DM.

In Ref.~\cite{FloquetLMG} nondissipative DTCs were predicted to exist in the $\gamma = 0$ LMG model under a kicking protocol, where the nonergodicity of the system was exploited to produce a period-doubling response in the spin component perpendicular to the $z$ direction. In this paper we present an alternative method of producing DTCs in the LMG model for arbitrary values of $\gamma$. Specifically, under a sinusoidally driven interaction parameter
\begin{equation}\label{eq:lambda_t}
    \lambda(t) = \lambda_0 [ 1 + A \sin (\omega_d t) ],
\end{equation}
we search for a clean period-doubling response in the dynamics of $\langle \hat{J}_x \rangle$, the order parameter of our system. In relation to this, Ref.~\cite{AC-LMG} explored the appearance of dynamical phases in a closed LMG model for arbitrary $\gamma$ by driving its interaction parameter sinusoidally. In this work we study via a semiclassical approach the previously unexplored effects of dissipation and thus consider the open LMG model in the context of dynamical phases that break time-translation symmetry. In addition, we will also check the robustness of the DTCs against changes in the initial conditions.

To begin our discussion of the open LMG model, we first present a generalization of its second-order quantum phase transitions, due to a broken $\mathbb{Z}_2$ symmetry in the system, for arbitrary values of $\gamma$. Recall that the $\mathbb{Z}_2$ symmetry of the DM can always be broken under certain combinations of system parameters. This, however, is not the case for the isotropic LMG model, as we present in the following section.

\subsection{Global Symmetry Breaking}\label{sec:IIIA}

\begin{figure}[t!]
    \centering
    \includegraphics[width=0.49\textwidth]{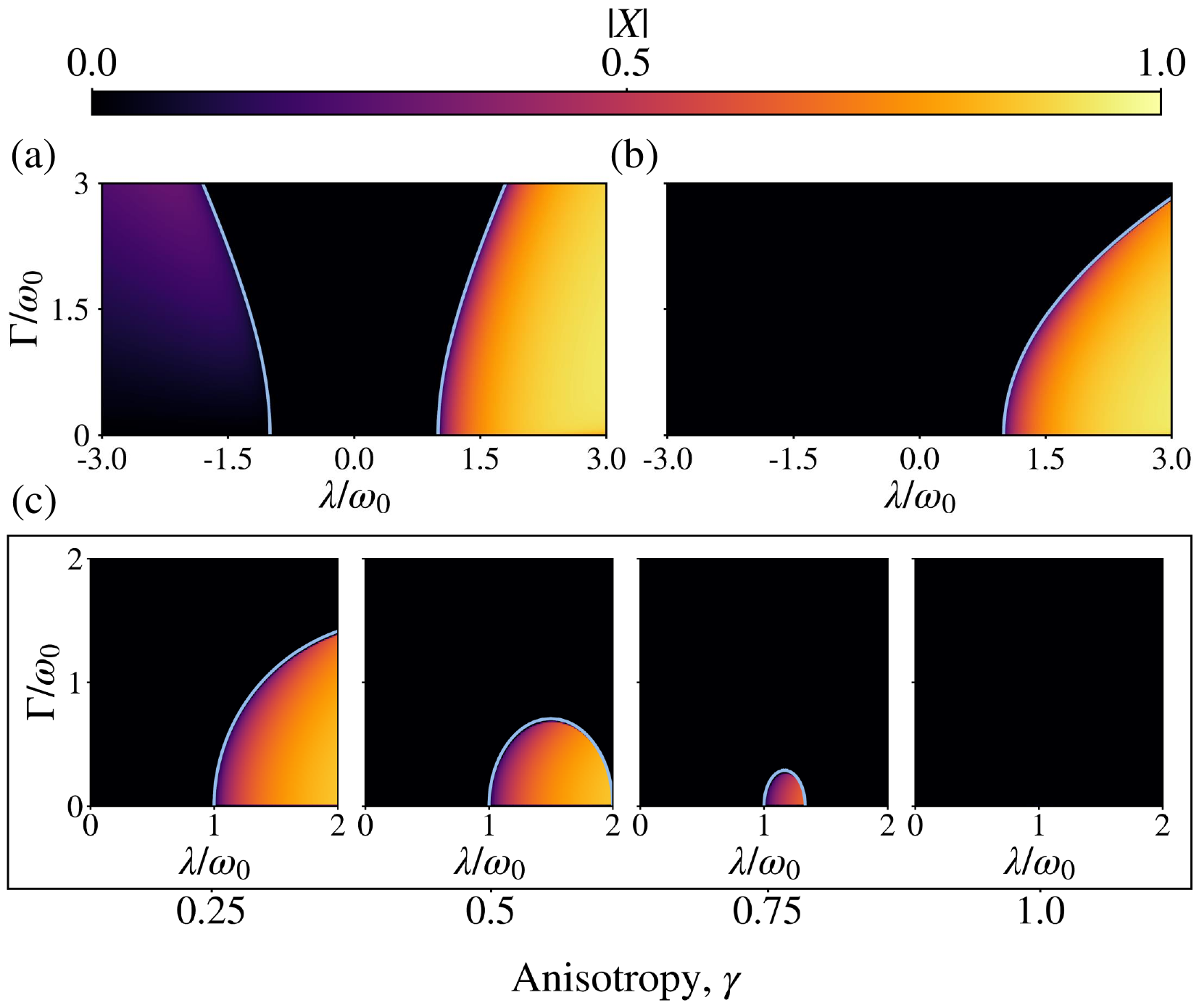}
    \caption{Global symmetry breaking in the LMG model in the (a) $\gamma = -1$, (b) $\gamma = 0$, and (c) $\gamma > 0$ regimes, as presented in the $\Gamma/\omega_0$ vs $\lambda/\omega_0$ space. The black region corresponds to the NP characterized by a steady-state solution of $|X| = 0$. The colored region corresponds to the SB phase with steady-state solution of $|X| \neq 0$. The blue curve represents the critical coupling strength $\lambda_c$.}
    \label{fig:undriven_pd}
\end{figure}

The generalized expression of the semiclassical equations of motion of the LMG model can be written as
\begin{subequations}\label{eq:LMG_eom}%
\begin{align}\label{eq:xdot}
    \dot{X} &= -\omega_0 Y - \lambda \gamma YZ + \frac{\Gamma}{2} XZ, \\
    \dot{Y} &= \omega_0X + \lambda XZ + \frac{\Gamma}{2} YZ, \\
    \dot{Z} &= -\lambda XY + \lambda \gamma XY - \frac{\Gamma}{2} (X^2 + Y^2), \label{eq:zdot}
\end{align}
\end{subequations}
which is obtained by introducing the dissipation parameter $\Gamma >0$, related to the collective spontaneous decay of a spin-up state to a spin-down state. We use the notation $X = 2\langle \hat{J}_x \rangle/N$, $Y = 2\langle \hat{J}_y \rangle/N$, and $Z = 2\langle \hat{J}_z \rangle/N$ to simplify the spin conservation equation
\begin{equation}
    X^2 + Y^2 + Z^2 = 1
\end{equation}
due to $[\hat{H}, \hat{J}^2] = 0$. A detailed derivation of Eq.~\eqref{eq:LMG_eom} is presented in Appendix~\ref{sec:LMG_eom}.

Analytically, one can obtain two types of steady-state solutions to Eq.~\eqref{eq:LMG_eom}, the first being the trivial $X_s = Y_s = 0$ and $Z_s = -1$, which corresponds to the NP. The second solution, corresponding to the SB phase, is given by
\begin{subequations}
    \begin{equation}\label{eq:X_s}
        X_s = \pm \frac{\Gamma}{2\Lambda} \sqrt{\frac{\Lambda^2 - \omega_0^2}{\lambda \gamma_- (\lambda - \Lambda)}},
    \end{equation}
    \begin{equation}
        Y_s = \frac{2(\Lambda -\lambda)}{\Gamma} X_s,
    \end{equation}
    and
    \begin{equation}\label{eq:LMG_zs}
        Z_s = \frac{-\omega_0}{\Lambda},
    \end{equation}
\end{subequations}
where $\Lambda = \bigl( \lambda \gamma_+ + \sqrt{\lambda^2 \gamma_-^2 -\Gamma^2} \bigr)/2$ and $\gamma_{\pm} = 1 \pm \gamma$. The critical coupling strength can be defined as the point where the system transitions between the two phases. This curve can be calculated analytically by setting Eq.~\eqref{eq:LMG_zs} equal to $-1$ and solving for $\lambda_c$ as
\begin{equation}\label{eq:l_crit}
    \lambda_c = \begin{cases}
        \frac{\Gamma^2}{4\omega_0} + \omega_0 \quad& \text{for } \gamma = 0 \\
        \frac{\pm \sqrt{\omega_0^2\gamma_-^2 - \Gamma^2 \gamma} \, + \, \omega_0\gamma_+}{2\gamma} \quad& \text{for } \gamma \neq 0.
    \end{cases}
\end{equation}

In  Fig.~\ref{fig:undriven_pd} we can observe these phase transitions for different values of $\gamma$. The initial states are set to $X_0 = 0, Y_0 = 5 \times 10^{-8},$ and $Z_0 = \sqrt{1 - Y_0^2}$, which are near but not equal to the lowest-energy state of the system. For the remainder of the paper, we use this set of initial states for all simulations unless specified otherwise. The analytical $\lambda_c$ are marked as the light solid curves on these plots, cleanly separating the normal and SB phase regions. Here we note that the existence of a second-order quantum phase transition in the positive and negative regimes of $\lambda/\omega_0$ only occurs when $\gamma < 0$. Observe that the SB phase region is largest when $\gamma = -1$, which eventually decreases in size as the anisotropy value is increased. In the $\gamma = 1$ limit, the SB phase region disappears, indicating the absence of $\mathbb{Z}_2$-symmetry breaking.

\subsection{Effective Oscillator Picture}

\begin{figure}[t!]
    \centering
    \includegraphics[width=0.49\textwidth]{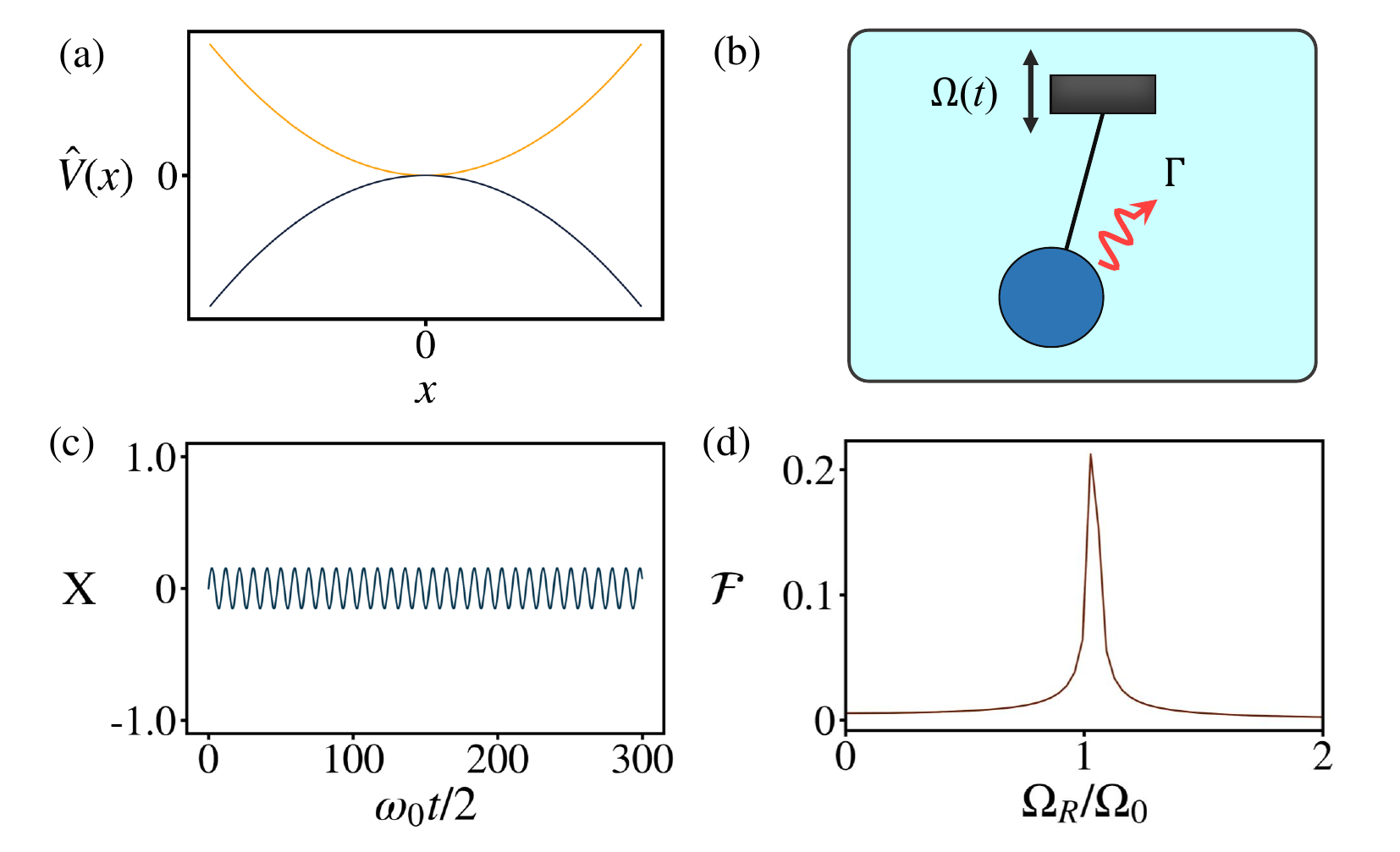}
    \caption{(a) Effective potential of the LMG Hamiltonian in pseudo-position-momentum representation. The potential opening upward (downward) is analogous to the system in the NP (SB phase). (b) Sketch of effective oscillator picture of the driven LMG model. (c) Dynamics of the undriven LMG model at $\{ \lambda/\omega_0, \Gamma/\omega_0, \gamma \} = \{ 0.9, 0, 0 \}$ with its (d) corresponding Fourier decomposition, where $\Omega_R$ is the response frequency. We have set $X_0 = 0$ and $Y_0 = 5\times 10^{-2}$ to emphasize the oscillations in $X$.}
    \label{fig:osc_LMG}
\end{figure}

One way to explain the conservation of $\mathbb{Z}_2$ symmetry in the isotropic LMG model is by mapping Eq.~\eqref{eq:LMG} into the pseudo-position-momentum representation. By applying a Holstein-Primakoff transformation in the $N \to \infty$ limit $(\hat{F} \approx 1)$ 
\begin{equation}
    \hat{J}_z = \hat{c}^{\dagger} \hat{c} - \frac{N}{2}, \qquad \hat{J}_- = \hat{J}_+^{\dagger} \cong \sqrt{N} \hat{c},
\end{equation}
one can express the LMG Hamiltonian in the bosonic picture via
\begin{equation}\label{eq:LMG_HP}
\begin{aligned}
    \frac{\hat{H}_{\text{HP}}}{\hbar} = &-\frac{\lambda}{4} \Bigl( 1 - \gamma \Bigr) \Bigl[ \bigl( \hat{c}^{\dagger} \bigr)^2 + \bigl( \hat{c} \bigr)^2 \Bigr] + \Bigl[ \omega_0 \\
    &- \frac{\lambda}{2} ( 1 + \gamma) \Bigr] \hat{c}^{\dagger} \hat{c} - \Bigl[ \frac{\omega_0 N}{2} + \frac{\lambda}{4}(1 + \gamma) \Bigr].
\end{aligned}
\end{equation}
We then express these bosonic operators in terms of pseudo-position and pseudo-momentum operators, i.e.,
\begin{subequations}\label{eq:HP_LMG}
    \begin{align}
        \hat{c}^{\dagger} &= \sqrt{\frac{\omega_0 - \lambda \gamma}{2}} \biggl( \hat{x} - \frac{i}{\omega_0 - \lambda \gamma} \hat{p} \biggr), \\
        \hat{c} &= \sqrt{\frac{\omega_0 - \lambda \gamma}{2}} \biggl( \hat{x} + \frac{i}{\omega_0 - \lambda \gamma} \hat{p} \biggr),
    \end{align}
\end{subequations}
leading us to an oscillator-like picture of the closed LMG model described by the Hamiltonian
\begin{equation}
    \frac{\hat{H}_{xp}}{\hbar} = \frac{1}{2} \biggl[ (\omega_0 - \gamma\lambda)(\omega_0 - \lambda) \hat{x}^2 + \hat{p}^2 - \omega_0(N+1) \biggr].
\end{equation}
An interpretation of this representation is shown in Fig.~\ref{fig:osc_LMG}(a). For $(\omega_0 - \gamma\lambda)(\omega_0 - \lambda) > 0$, we obtain an effective parabolic potential that only admits bound states, equivalent to the system stabilizing to the NP. However, when $(\omega_0 - \gamma\lambda)(\omega_0 - \lambda) < 0$, this potential opens downward and now forces the system to pick one of two possible configurations, i.e., $\mathbb{Z}_2$-symmetry breaking. This latter condition can be satisfied for certain combinations of $\omega_0$ and $\lambda$ for all but the $\gamma = 1$ case, in which $(\omega_0 - \lambda)^2 \nless 0$.

\begin{figure*}[t!]
    \centering
    \includegraphics[width=\textwidth]{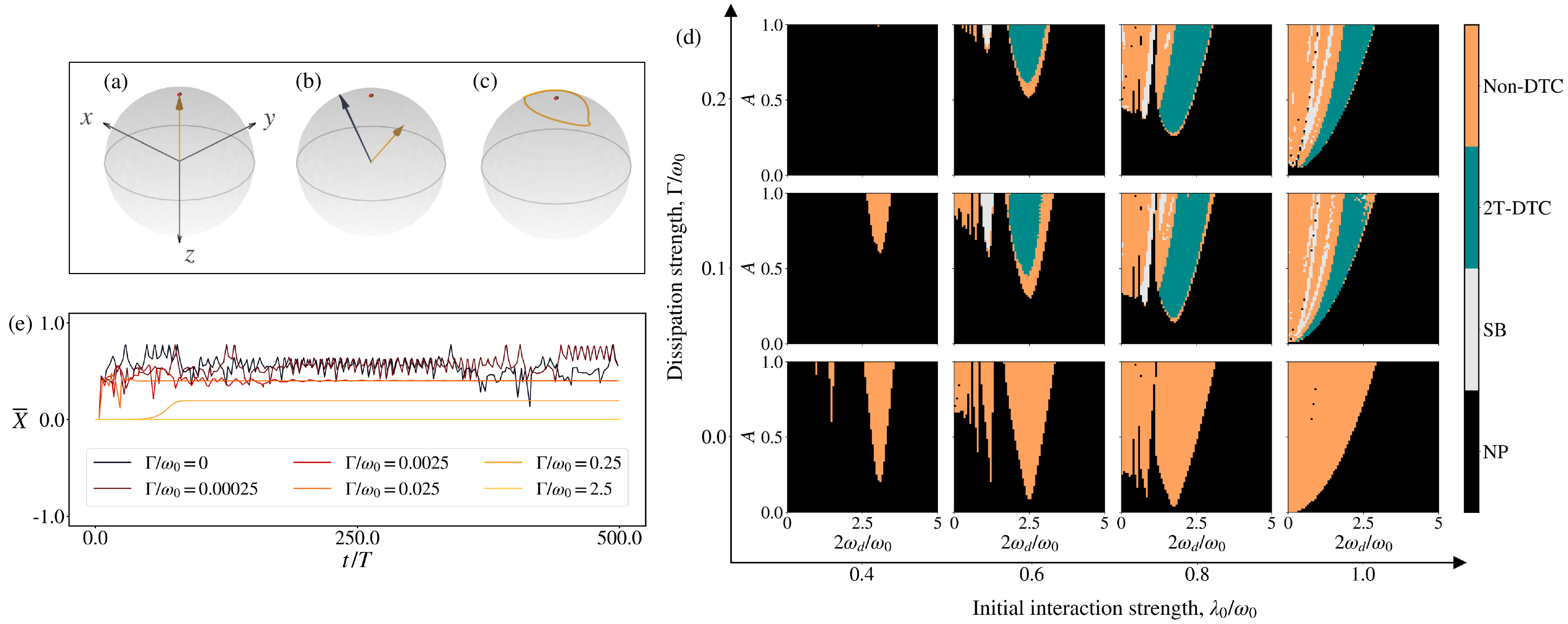}
    \caption{Bloch sphere dynamics of (a) NP, (b) SB, and (c) $2T$-DTC phases, respectively. The blue (yellow) trajectories mark the dynamics due to an initial state marked by the green (red) dots. These dynamics are observed in the parameter regions (a) $\{ \lambda/\omega_0 , \Gamma/\omega_0 \} = \{ 0.75, 0.03 \}$, (b) $\{ \lambda/\omega_0 , \Gamma/\omega_0 \} = \{ 1.25, 0.03 \}$, and (c) $\{ \lambda_0/\omega_0 , \Gamma/\omega_0, A, 2\omega_d/\omega_0 \} = \{ 0.8, 0.1, 0.5, 1.7 \}$. (d) Phase diagrams of the driven LMG model at varying dissipation and initial interaction strengths. (e) Oscillation envelopes of $X$ at varying dissipation strengths for $\{ \lambda_0/\omega_0 , A, 2\omega_d/\omega_0 \} = \{ 0.95, 0.15, 0.9 \}$. All results are for the $\gamma = 0$ case.}
    \label{fig:LMG_pd}
\end{figure*}

As a precursor to the discussion in the following section, we also derive the parametric oscillator picture of the LMG model. This can be obtained by calculating the equations of motion of the LMG model in the $x$-$p$ representation under a parametric drive given by Eq.~\eqref{eq:lambda_t}. Using Eq.~\eqref{eq:LMG_HP} and $\langle \hat{c} \rangle = c$, one obtains the time evolution of the bosonic operators as
\begin{equation}
    \dot{c} = \frac{i}{2} \lambda(t)(1 - \gamma) c^* - i \biggl[ \omega_0 - \frac{\lambda(t)}{2}(1 + \gamma) \biggr] c - \frac{\Gamma}{2} c.
\end{equation}
Following Eq.~\eqref{eq:HP_LMG}, this can be mapped to the pseudo-position representation as
\begin{equation}\label{eq:34}
    \ddot{x} + \Gamma \dot{x} + \{ [ \omega_0 - \lambda(t) ][\omega_0 - \gamma \lambda(t)]\} x = 0,
\end{equation}
with $\langle \hat{x} \rangle = x$. By introducing the variable $r = e^{\Gamma t} x$, we are then able to rewrite Eq.~\eqref{eq:34} as
\begin{equation}
    \ddot{r} + \Omega^2(t) r = 0,
\end{equation}
with $\Omega^2(t) = \Omega_0^2 (1 + f(t))$, where
\begin{equation}
    \Omega_0 = \sqrt{(\omega_0 - \lambda_0)(\omega_0 - \gamma \lambda_0)}
\end{equation}
is the natural frequency of the system and $f(t)$ is the time-dependent part of the equation. This natural frequency can be observed at sufficiently large $X_0$ or $Y_0$ and at $\Gamma/\omega_0 = 0$, where the long-time dynamics of $X$, $Y$, and $Z$ do not decay over time. Note that this treatment is only valid when $\Omega_0 \in \mathbb{R}$; as such, for the remainder of this section, we will only consider the parameter regime $0 \leq \lambda_0/\omega_0 \leq 1$. A schematic of the LMG model in the parametric oscillator picture is presented in Fig.~\ref{fig:osc_LMG}(b), while the sample dynamics and Fourier spectrum displaying the natural frequency of the LMG model are shown in Figs.~\ref{fig:osc_LMG}(c) and \ref{fig:osc_LMG}(d), respectively. These arguments imply that parametric resonance theory can also be used to describe time crystal formation in the open LMG model, similar to the DM from the previous section.

\subsection{Time Translation Symmetry Breaking} \label{sec:LMG_TTSB}

By introducing a periodic drive in the system given by Eq.~\eqref{eq:lambda_t}, additional dynamical phases can be accessed by the LMG model at different combination of system parameters. In Figs.~\ref{fig:LMG_pd}(a)--\ref{fig:LMG_pd}(c) we present the Bloch sphere dynamics of one of these phases, the $2T$-DTC phase, in a time window $t_0 < t < t_0 + 2T$, where $T$ is the driving period, along with the steady-state dynamics of the NP and SB phases mentioned in Sec.~\ref{sec:IIIA}. The behavior of other observed dynamical (non-DTC) phases are also discussed in Appendix~\ref{sec:nondtc_lmg}. The differences in these phases can be clearly observed in this representation. Figures~\ref{fig:LMG_pd}(a) and \ref{fig:LMG_pd}(b) present the NP and SB phase steady states, where the former corresponds to the collective spin relaxing to the spin-down configuration, i.e., the lowest-energy state of Eq.~\eqref{eq:LMG}. The latter is shown to have two possible steady states for different initial conditions, which is due to $X_s$ (and consequently $Y_s$) having both positive and negative values. This behavior corresponds to the broken $\mathbb{Z}_2$ symmetry of the system as $X$ is forced to pick one of the two possible steady-state values, indicated by $\pm$ in Eq.~\eqref{eq:X_s}. Figure~\ref{fig:LMG_pd}(c), on the other hand, shows the dynamics of the $2T$-DTC phase, where the oscillation frequencies of $X$ peak at half the driving frequency $\Omega_r = \omega_d/2$, and has a constant oscillation amplitude. We also observe the presence of secondary peaks in the frequency spectrum of the $2T$-DTC, specifically at harmonic odd multiples of $\omega_d/2$ similar to Ref.~\cite{Ojeda_2023}.

To show a better picture of when these phases appear in the driven open LMG model, we map the dynamics of the system onto phase diagrams as functions of the driving amplitude $A$ and driving frequency $\omega_d$, for different combinations of system parameters. We present in Fig.~\ref{fig:LMG_pd}(d) a set of phase diagrams of the anisotropic ($\gamma = 0$) LMG model for varying initial interaction and dissipation strengths. Notice that $2T$-DTC phases exist only when dissipation is introduced in the system, similar to the DM in the previous section. To show how $\Gamma$ affects the dynamics of the system, we plot in Fig.~\ref{fig:LMG_pd}(e) the oscillation envelopes $\overline{X}$ for varying dissipation strengths, keeping all other system parameters constant. When there is virtually no dissipation (e.g., $\Gamma/\omega_0 = 0$ or $2.5 \times 10^{-4}$), the system thermalizes into a non-DTC phase as the amount of energy going into the system continually builds up over time. By introducing a small dissipation channel $\Gamma/\omega_0 = (2.5 \times 10^{-3})$-$(2.5 \times 10^{-2})$, the system is allowed to relax to the $2T$-DTC phase. However, by further increasing the dissipation strength ($\Gamma/\omega_0 = 0.25$), the system then requires a longer time to build up energy to access this phase, as can be seen by the slower increase in the amplitude of $X$. Eventually, when dissipation becomes too strong ($\Gamma/\omega_0 = 2.5$), the system becomes unable to gain enough energy to access any dynamical phase, therefore stabilizing to the NP.

From the preceeding section, the natural frequency of the system is dependent on the initial interaction strength $\lambda_0$ and hence so are the resonance frequencies defined by $\omega_r = 2\Omega_0/n$, where $n \in \mathbb{N}$. Each value of $n$ corresponds to a different resonance lobe frequency, with $n = 1$ the primary resonance lobe in the phase diagrams. In the LMG model, DTCs exist exclusively in the primary resonance lobe. At $n > 1$, the system can only access non-DTC phases. Also note how the instability region decreases in size as $\lambda_0$ decreases. This can be attributed to the interaction strength being too far away from $\lambda_c$, and therefore requiring a larger minimum driving amplitude to break $\mathbb{Z}_2$ symmetry.

\subsection{The Isotropic LMG Model} \label{sec:LMG_limit}
\begin{figure}[t!]
    \centering
    \includegraphics[width=0.49\textwidth]{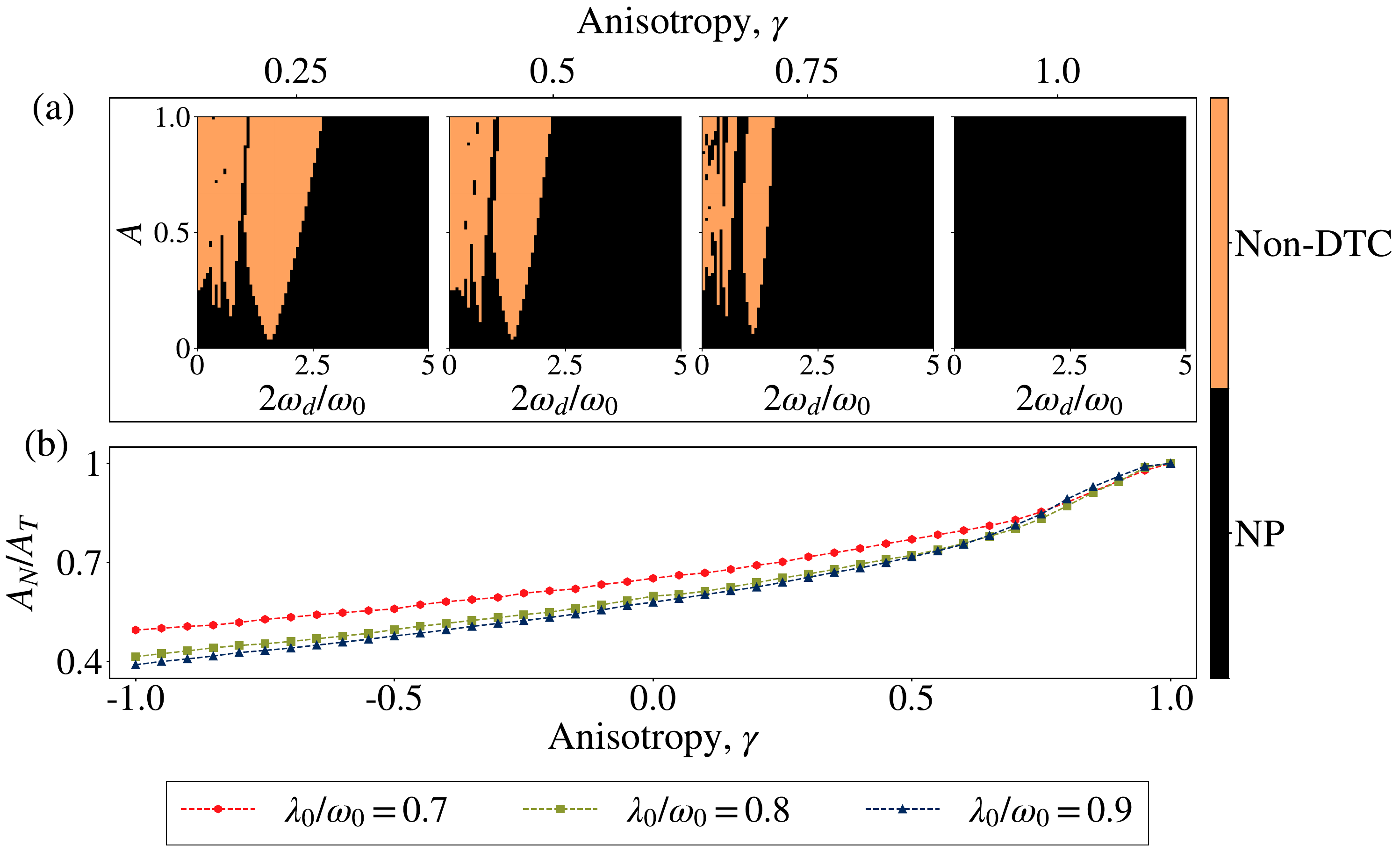}
    \caption{(a) Phase diagrams for the driven LMG model at varying anisotropy value, where $\Gamma = 0$ and $\lambda_0/\omega_0 = 0.8$. Note how the instability region decreases in size as $\gamma$ increases, until eventually disappearing when $\gamma = 1$. This behavior is further proven by (b) $A_N/A_T$, the ratio between the area of the NP region and the total area of each phase diagram, where we can see that for different cases of $\lambda_0$, the resonance lobe always disappears eventually as $\gamma \to 1$.}
    \label{fig:LMG_gamma}
\end{figure}

To conclude this analysis on the LMG model, we consider what happens when we vary the anisotropy parameter of the driven LMG model. In Fig.~\ref{fig:LMG_gamma}(a) we present phase diagrams at different values of $\gamma$ in the $\Gamma = 0$ case. Notice how the width of the resonance lobes decreases as the anisotropy parameter increases from $\gamma = 0.25$ to $\gamma = 1$. To quantify this observation, we define $A_N/A_T$ as the ratio between the NP region and the total phase diagram area such that $A_N/A_T = 1$ implies that the NP dominates the phase diagram and thus the system does not have a DTC phase. We demonstrate in Fig.~\ref{fig:LMG_gamma}(b) more clearly that $A_{N}/A_{T} \rightarrow 1$ as $\gamma \rightarrow 1$, indicating that the resonance lobes always disappear in the isotropic LMG limit. Note that the measures of $A_N$ and $A_T$ are limited to the phase diagram areas where $A \in [0, 1]$ and $2\omega_d/\omega_0 \in [0, 5]$.

Relating Figs.~\ref{fig:undriven_pd} and \ref{fig:LMG_gamma}, we see that in both the driven and undriven cases, the SB (instability) regions are largest at $\gamma = -1$, which gradually decrease in size as $\gamma$ increases. Phase transitions are most accessible as $\gamma \to -1$ and become less prominent as $\gamma \to 1$. In the special case of $\gamma = 1$, the resonance lobes disappear completely, analogous to the conservation of $\mathbb{Z}_2$ symmetry in the undriven case. This shows that in fully connected spins a global symmetry, in this case the $\mathbb{Z}_2$ symmetry, is a prerequisite for time-translation symmetry breaking, as pointed out in other systems in Refs.~\cite{Khemani_PRE, Khemani_PRL, FloquetLMG, II_von, Pizzi, Floquet}.

To further elaborate on this claim, let us consider the solutions of Eq.~\eqref{eq:LMG_eom} when $\lambda$ is given by Eq.~\eqref{eq:lambda_t}. In the nondissipative $\gamma = 1$ case, we have $\dot{Z}=0$ and therefore $Z(t) = Z_0$. Exploiting spin conservation, the solution to Eq.~\eqref{eq:xdot} is a sinusoidal function
\begin{equation}
    \begin{aligned}
    X(t) = &\sqrt{1 - Z_0} \sin \biggl\{ -(\omega_0 + \lambda_0 Z_0) t + \frac{\lambda_0 Z_0 A}{\omega_d} \\
    &\times \Bigl[ \cos (\omega_d t) - 1 \Bigr] + \sin^{-1} \biggl( \frac{X_0}{\sqrt{1 - Z_0}} \biggr) \biggr\}.
    \end{aligned}
\end{equation}
This means that the dynamics of $X$ oscillates about $X = 0$ over time without decay, its amplitude being dependent only on $Z_0$. Signatures of parametric resonance require an initially exponential increase in the order parameter \cite{kovacic_mathieus_2018}, which does not occur in this case. Also, since we have considered $Z_0 \approx 1$, the oscillation amplitude would be very small and can effectively be taken as zero; thus the nondissipative case can only host the NP. In the presence of dissipation, Eq.~\eqref{eq:zdot} depends only on $\Gamma$ and has the solution
\begin{equation}
    Z(t) = \frac{e^{\Gamma t} - \frac{1 - Z_0}{1 + Z_0}}{e^{\Gamma t} + \frac{1 - Z_0}{1 + Z_0}}.
\end{equation}
In the $t \to \infty$ limit, $Z(t) \to -1$. Therefore, via spin conservation, the steady state of $X$ is $0$, and the system relaxes to the NP. This implies that only the NP exists in the $\gamma = 1$ case of the LMG model.

\section{Summary and Discussion}\label{sec:conclusion}

In this work we proposed a theory based on parametric resonance as the main mechanism for the formation of discrete time crystals in fully connected spin-cavity systems. We examined the role of nonlinearity and dissipation through the lens of parametric instability by mapping the open DM into a system of dissipative coupled oscillator models. Symmetry breaking was analyzed through the mean-field dynamics of the open LMG model with tunable anisotropy.

We have shown in the dynamics of the OM that nonlinearity prevents the system's dynamics from unbounded exponential growth, keeping the system's dynamics physical. Meanwhile, the dissipation allows the system to relax into a clean period-doubling response in finite time, which is a key feature of the DTCs observed in experimental setups of driven-dissipative systems \cite{kesler_observation_2021, kongkhambut_realization_2021, heugel_classical_2019, frey_realization_2022}. This constitutes an alternative method for stabilizing DTCs in open systems beyond the many-body localization used for closed systems \cite{Khemani_PRE, Floquet, II_von}. We have also demonstrated that while dissipation is an important ingredient in DTC formation, it can modify the system's phase diagram to a certain degree depending on its strength. In particular, the primary resonance corresponding to the largest DTC region increases with $\kappa$ until it saturates at a larger value given by $\Omega_{\kappa \gg \omega}$. The minimum driving amplitude needed to enter the DTC phase is also largest at $\kappa = \omega$, implying that the parametric instability associated with DTCs is only favorable for either small or large dissipation strength. Finally, we have observed that the non-DTC regions increase in size as $\kappa \rightarrow \infty$. We have attributed this behavior to the open NOM becoming an effective single oscillator with negligible dissipation in the bad cavity limit due to the degeneracy of the polariton modes and the suppression of its two eigenmodes by the dissipation. This raises the question of the ultimate fate of DTCs in the large-$\kappa$ regime, where the system experiences an effectively weak dissipation despite $\kappa$ being significantly larger than the values considered in this work.

We have also demonstrated the role of $\mathbb{Z}_{2}$-symmetry breaking in the DTC formation by analyzing the dynamics of the periodically driven LMG model. In this system, the nonlinearity comes from the anisotropic spin interaction, while the dissipation comes from the global spin decay. Similar to the open DM, we have only observed DTCs in the LMG model when dissipation is present. Moreover, the dynamical phase only appears when the system has a symmetry-broken phase in the undriven case, which happens only when the spin interaction is not isotropic. Without anisotropy, the conservation of the collective spin's angular momentum and its $z$ component prevents parametric instability from appearing in the closed limit, while the dissipation forces the system to relax into its steady state in the long-time limit. Thus, there is no stable DTC phase in the isotropic limit of the LMG model. Note that this scenario does not occur in the DM since it always has a symmetry-broken phase in the form of the superradiant phase, regardless of the value of $\omega$ and $\omega_{0}$.

Given the tunability of the symmetry breaking in the LMG model, a natural extension of our work would be to explore how dynamically changing the anisotropy parameter affects the conditions necessary for the existence of DTCs. This question can also be extended to systems with tunable symmetries, such as the general Dicke model \cite{bhaseen_dynamics_2012}. Furthermore, it would be interesting to explore the nature and stability of the various types of transient dynamical response that we categorized here as non-DTC using methods \cite{nicolaou_complex_2024} better suited for identifying attractors in dynamical systems.

\section*{Acknowledgement}
This work was funded by the UP System Balik Ph.D. Program through Grant No. OVPAA-BPhD-2021-04.

\appendix

\section{Breakdown of Oscillator Models}\label{sec:appendix_breakdown}

In deriving the Hamiltonian of the effective oscillator models of the DM, we implicitly assumed that the atomic excitation number $\hat{b}^{\dagger}\hat{b}$ is smaller than $N$, i.e. $\hat{b}^{\dagger}\hat{b} / N \ll 1$, throughout its dynamics to truncate $\hat{F}$. The oscillator models break down the moment that $|b|^{2}/N > 1$ for any time $t$. Here we will demonstrate that we can relax the assumption by showing that the deviation of $J_{x} / N \approx \Re(b) / \sqrt{N}$ from its true value remains small when we consider the dynamics of the NOM.

\begin{figure}[t!]
\centering
    \includegraphics[scale = 0.4]{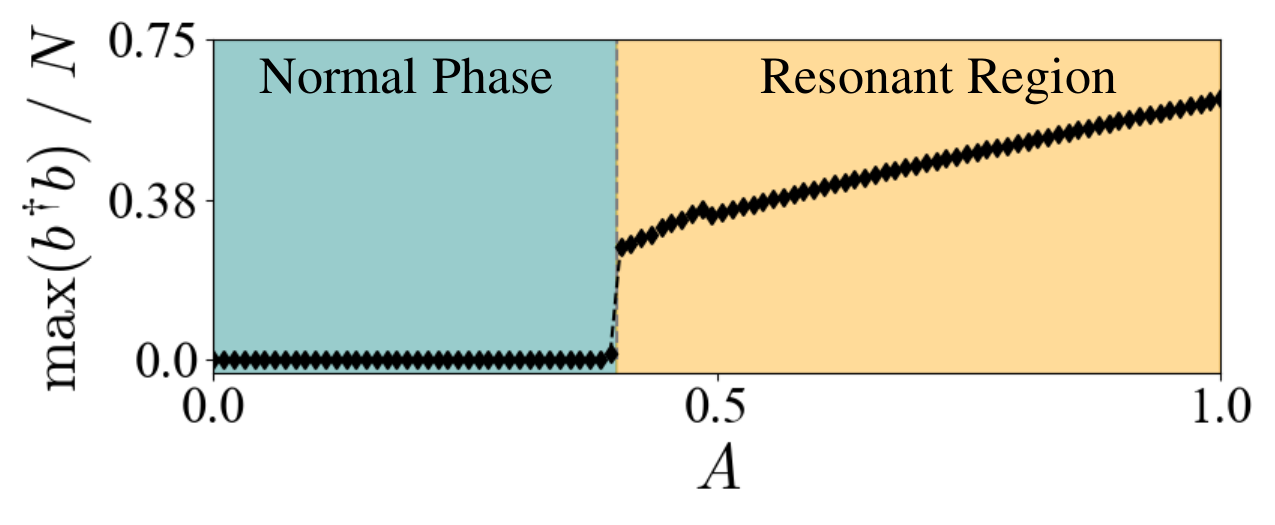}
    \caption{ Maximum response amplitude of $|b|^{2}/N$ as a function of $A$, when the system is driven close to resonance. The remaining parameters are set to $\kappa = 0.5 \omega$, $g_{0} = 0.9 g_{c}$, and $\omega_{d} = 0.8 \omega$. }
    \label{fig:om_breakdown}
\end{figure}

To start, let us consider the Taylor expansion of $J_{x}$ in the mean-field limit, 
\begin{equation}
\label{eq:jx_meanfield}
\begin{aligned}
J_{x} &= \sqrt{N} \Re(b) \sqrt{1 - \frac{|b|^{2}}{N}} \\
      &\approx \sqrt{N} \Re(b) \left( 1 - \frac{1}{2} \frac{|b|^{2}}{N} - \frac{1}{8} \left(\frac{|b|^{2}}{N}\right)^{2} + \ldots  \right) .
\end{aligned}
\end{equation}
If we subtract the approximation of $J_{x}$ in the oscillator level $J_{x} \approx \sqrt{N} \Re(b)$ in Eq.~\eqref{eq:jx_meanfield}, we see that the deviation is
\begin{equation}
\label{eq:spin_deviation}
\frac{\triangle J_{x}}{N} = \frac{J_{x}}{N} - \frac{\Re(b)}{\sqrt{N}} \sim O\left( \left( \frac{|b|}{\sqrt{N}} \right)^{3} \right),
\end{equation} 
that is, the deviation of $\Re(b)/\sqrt{N}$ from $J_{x}/N$ is of order $(|b|/\sqrt{N})^{3}$, which remains small so long as $|b|/\sqrt{N} < 1$. As shown in Figs.~\ref{fig:main_results}(c) and \ref{fig:main_results}(d), this condition does not hold for the driven LOM since $|b| \rightarrow \infty$ as $t \rightarrow \infty$. As for the NOM, the dynamics of $\Re(b)$ is bounded, which is why the model remains applicable for driving parameters that do not lead to dynamics with $|b|/ \sqrt{N} > 1$. We demonstrate in Fig.~\ref{fig:om_breakdown} that the maximum response amplitude of $|b|^{2}/N $ remains less than one within the interval $A = [0, 1]$ when the system is driven close to resonance. This result confirms that the NOM remains a good approximation of the DM for all driving amplitudes considered in the main text.

\section{OM Equations of Motions}
\label{sec:om_eom}

We start our derivation of the equations of motions for the oscillator models with the master equation of the open LOM for the expectation value of an arbitrary operator, $\hat{\mathcal{O}}$,
\begin{equation}
\label{eq:expectvalue_mastereq_DM}
\partial_{t}\left< \hat{\mathcal{O}} \right> = \left<  i \left[\frac{\hat{H}_{\mathrm{LOM}}}{\hbar}, \hat{\mathcal{O}} \right] + \kappa D[\hat{a}]\hat{\mathcal{O}} \right>,
\end{equation}
where $D[\hat{a}]\hat{\mathcal{O}} = 2\hat{a}^{\dagger} \hat{\mathcal{O}} \hat{a} - \left\{ \hat{a}^{\dagger}\hat{a}, \hat{\mathcal{O}} \right\}$. The equations of motions of $a = \left< \hat{a} \right>$ and $b = \left< b \right>$ under the mean-field approximation is given as
\begin{subequations}
\label{eq:lom_atom_cavity_eom}
	\begin{equation}
	i\partial_{t} a = \omega a + g \left( b + b^{*} \right) - i \kappa a,
	\end{equation}
	\begin{equation}
	i \partial_{t} b = \omega_{0} b + g \left( a + a^{*} \right),
	\end{equation}
\end{subequations}
following the commutator relations $\left[ \hat{a}, \hat{a}^{\dagger} \right] = \left[\hat{b} , \hat{b}^{\dagger} \right] = 1$. To turn Eq.~\eqref{eq:lom_atom_cavity_eom} into the known form for coupled oscillators, we substitute the pseudoposition and pseudomomentum representations for the cavity mode 
\begin{equation}
x = \frac{1}{\sqrt{2 \omega}} \left( a + a^{*} \right), \quad p_{x} = i \sqrt{\frac{\omega}{2}} \left( a - a^{*} \right)
\end{equation}
and the atomic mode
\begin{equation}
y = \frac{1}{\sqrt{2 \omega_{0}}} \left( b + b^{*} \right), \quad p_{x} = i \sqrt{\frac{\omega_{0}}{2}} \left( b - b^{*} \right)
\end{equation}
into Eq.~\eqref{eq:lom_atom_cavity_eom} to get a pseudo-Hamilton equation of the form 
\begin{subequations}
\label{eq:lom_pseudo_hamilton}
	\begin{equation}
	\dot{x} = p_{x} - \kappa x, \quad \dot{p}_{x} = - \omega^{2}x - 2 g \sqrt{\omega \omega_{0}} y - \kappa p_{x}, 
	\end{equation}
 	
	\begin{equation}
	\dot{y} = p_{y}, \quad \dot{p}_{y} = - \omega_{0}^{2}y - 2 g \sqrt{\omega \omega_{0}} x .
	\end{equation}	 	
\end{subequations}
Solving for $x$ and $y$ and setting $\omega = \omega_{0}$ gives us the second-order differential equations \eqref{eq:lom_in_xy}.

We can also derive the equations of motions for the NOM and its corresponding pseudo-Hamilton equation by substituting $\hat{H}_{\mathrm{NOM}}$ into Eq.~\eqref{eq:expectvalue_mastereq_DM}. This leads to a nonlinear mean-field equation of the form
\begin{subequations}
\label{eq:nom_atom_cavity_eom}
	\begin{equation}
        \label{eq:nom_cavity_eom}
	i\partial_{t} a = \omega a + g \left( b + b^{*} \right) \left( 1 - \frac{|b|^{2}}{2N} \right) - i \kappa a,
	\end{equation}
	\begin{equation}
        \label{eq:nom_atom_eom}
	i \partial_{t} b = \omega_{0} b + g \left( a + a^{*} \right)\left( 1 - \frac{1}{2N} \left( 2|b|^{2} + b^{2} \right) \right)  ,
	\end{equation}
\end{subequations}
where its equivalent pseudo-Hamilton equation is
\begin{subequations}
\label{eq:nom_pseudo_hamilton}
	\begin{equation}
	\dot{x} = p_{x} - \kappa x, \quad  \dot{y} = p_{y} - \frac{g}{N}xyp_{y}, 
	\end{equation}
        \begin{equation}
        \begin{aligned}
        \dot{p}_{x} &= - \omega^{2}x - 2 g \sqrt{\omega \omega_{0}} y - \kappa p_{x}   \\
         &  + \frac{g}{2N} \sqrt{\frac{\omega}{\omega_{0}}} \left( \omega_{0}^{2}y^{3} + yp_{y}^{2} \right),
        \end{aligned}
        \end{equation}

	\begin{equation}
        \begin{aligned}
        \dot{p}_{y} & = - \omega_{0}^{2}y - 2 g \sqrt{\omega \omega_{0}} x \\
        & + \frac{g}{2N} \sqrt{\frac{\omega}{\omega_{0}}} \left( 3\omega_{0}^{2}xy^{2} + xp_{y}^{2} \right) . 
        \end{aligned}
	\end{equation}	 	
\end{subequations}
To cast Eq.~\eqref{eq:nom_pseudo_hamilton} into a second-order differential equation, we assume that $N \rightarrow \infty$ so that $\dot{y} \approx p_{y}$, reducing Eq.~\eqref{eq:nom_pseudo_hamilton} to
\begin{subequations}
    \label{eq:nom_in_xy}
    \begin{equation}
        \ddot{x} + 2 \kappa \dot{x} = - \left( \omega^{2} + \kappa^{2} \right) x - 2 g \omega y + \frac{g}{2N}\left( \omega^{2}y^{3} + y\dot{y}^{2} \right), 
    \end{equation}
    \begin{equation}
        \ddot{y} = - 2 g \omega x - \omega^{2} y + \frac{g }{2N}\left( 3\omega^{2}xy^{2} + x\dot{y}^{2} \right).
    \end{equation}
\end{subequations}
where we again set $\omega = \omega_{0}$. In this form, it becomes more evident that the nonlinearity coming from the spin-cavity interaction corresponds to a Kerr-like nonlinear coupling between the two oscillators.

\section{Polariton Modes of the DM} \label{sec:om_resonance}

\subsection{Diagonalization of Open LOM}
\label{subsec:diagonalization}

To diagonalize Eq.~\eqref{eq:lom_in_xy}, it is convenient to write the equations of motions of $x$ and $y$ in matrix form 
\begin{equation}
\label{eq:lom_in_xy_matrix}
\partial_{t}^{2} \textbf{x} + 
\begin{pmatrix}
2\kappa & 0 \\
0 & 0
\end{pmatrix} \partial_{t}\textbf{x} + 
\begin{pmatrix}
\omega^{2} + \kappa^{2} & 2 g \omega \\
2 g\omega         & \omega^{2}
\end{pmatrix} 
\textbf{x} = 0,
\end{equation}
where $\textbf{x} = \left[ x \quad y \right]^{T}$ and $T$ represents transposition. In this form, we can substitute the ansatz
\begin{equation}
\label{eq:diagonal_ansatz}
\textbf{x} =  \frac{1}{2} \exp \left(- \frac{\kappa t}{2} \right)
\begin{pmatrix}
1 & 1\\
1 & -1
\end{pmatrix}
\textbf{x'},
\end{equation}
where $\textbf{x} = \left[ x_{+} \quad x_{-}\right]^{T}$, into Eq.~\eqref{eq:lom_in_xy_matrix} to reduce it to 
\begin{equation}
\label{eq:dissipative_coupling_om}
    \ddot{x}_{\pm} + \left( \omega^{2} + \frac{\kappa^{2}}{4} \pm 2g \omega \right)x_{\pm} = - \kappa \dot{x}_{\mp}.
\end{equation}
Note that in the closed limit $\kappa = 0$, the two oscillators in Eq.~\eqref{eq:dissipative_coupling_om} become uncoupled, leaving us with the two polariton modes of the closed DM resonantly driven at
\begin{equation}
\Omega_{\pm}(\kappa = 0) = \sqrt{\omega^{2} \pm 2g\omega}.
\end{equation}
This is consistent with the resonant frequencies derived from the closed DM in the Heisenberg picture \cite{bastidas_entanglement_2010, bastidas_nonequilibrium_2012}.

For nonzero $\kappa$, we can further diagonalize Eq.~\eqref{eq:dissipative_coupling_om} by converting it into a set of first-order differential equations. Introducing the variables $v_{\pm} = \dot{x}_{\pm}$ and the vector $\textbf{u} = \left[x_{+} \quad \dot{x}_{+} \quad x_{-} \quad \dot{x}_{-} \right]^{T}$, Eq.~\eqref{eq:dissipative_coupling_om} can be rewritten as
\begin{equation}
\partial_{t} \textbf{u} = \textbf{M} \textbf{u},
\end{equation}
where
\begin{equation}
\label{eq:x_pm_jacobian}
\textbf{M} = 
\begin{pmatrix}
0 & 1 & 0 & 0\\
-\omega_{+}^{2} & 0 & - \kappa & 0 \\
0 & 0 & 0 & 1\\
0 & -\kappa & -\omega_{-}^{2} & 0 
\end{pmatrix},
\end{equation}
with
\begin{equation}
\omega_{\pm}^{2} = \omega^{2} + \frac{\kappa^{2}}{4} \pm 2g \omega.
\end{equation}
In this form, diagonalizing Eq.~\eqref{eq:dissipative_coupling_om} is equivalent to diagonalizing $\textbf{M}$, i.e. finding the similarity transformation $\textbf{M} = \textbf{S} \textbf{D} \textbf{S}^{-1} $, where $\textbf{D}$ is a diagonal matrix containing the eigenvalues of $\textbf{M}$. Since we are only interested in finding the polariton modes of the open LOM, it suffices to determine $\textbf{D}$, which we found to be 
\begin{equation}
\textbf{D} = \mathrm{diag}\left( -i\Omega_{+}, i\Omega_{+}, -i\Omega_{-}, i \Omega_{-}  \right),
\end{equation}
where
\begin{equation}
\Omega_{\pm} = \left(\omega^{2} - \frac{\kappa^{2}}{4} \pm  2\omega \sqrt{g^{2} - \frac{\kappa^{2}}{4} }  \right)^{1/2} 
\end{equation}
are the polariton mode frequencies of the open LOM. For completion, let us note that if we apply the transformation 
\begin{equation}
\partial_{t}\textbf{u} = \textbf{S}\textbf{D}\textbf{S}^{-1} \textbf{u} \rightarrow \partial_{t}\textbf{u}' = \textbf{D} \textbf{u}',
\end{equation} 
and assume $\textbf{u}' = \textbf{S}^{-1}\textbf{u}= \left[ X_{+} \quad X_{+}^{*} \quad X_{-} \quad X_{-}^{*} \right]^{T}  $, we can derive Eq.~\eqref{eq:driven_lom_diagnoalized} by simply solving for the second-order differential equations of $X_{\pm}$.

\subsection{Eigenmodes of the Undriven Open LOM}
\label{subsec:eigenmodes}

From the forms of Eqs.~\eqref{eq:diagonal_ansatz} and \eqref{eq:driven_lom_diagnoalized} in the limit of $A = 0$, we can infer that the fundamental solutions of $x$ and $y$ are proportional to 
\begin{equation}
\textbf{x} \propto \exp\left( \varepsilon_{\pm}^{(\mathrm{U}, \mathrm{L})} t \right)\textbf{A},
\end{equation} 
where \textbf{A} is a constant vector and 
\begin{subequations}
\begin{equation}
\varepsilon_{\pm}^{\mathrm{U}} = - \frac{\kappa}{2} + i \Omega_{\pm},
\end{equation}
\begin{equation}
\varepsilon_{\pm}^{\mathrm{L}} = - \frac{\kappa}{2} - i \Omega_{\pm}
\end{equation}
\end{subequations}
are the upper and lower branches of the eigenmodes of the open LOM, respectively. This is consistent with the eigenvalues of the linearized open DM for the superradiant phase derived in Ref. \cite{dimer_proposed_2007}. Note that within the interval $\kappa_{c}' \leq \kappa_{c}''$, $\Omega_{-}$ becomes imaginary while the upper polariton frequency becomes imaginary at the interval $\kappa_{+}' \leq \kappa \leq \kappa_{c}''$, where
\begin{equation}
    \kappa_{+}' = 2\omega \left( 2g'^{2} - 1 + g'\sqrt{4g'^{2} - 3}  \right)^{1/2}.
\end{equation}
For the value of $g'$ considered in the main text, $\kappa_{+}' \approx \kappa_{c}''$; hence we only consider the behavior of the lower polariton mode as a function of $\kappa$.

For completeness, the real and imaginary components of the polariton frequencies at $\kappa > \kappa_{c}''$ are
\begin{equation}
\Re(\Omega_{\pm}) = \frac{1}{\sqrt{2}}\left[\frac{\kappa^{2}}{4} - \omega^{2} + \frac{\kappa^{2}}{4} \sqrt{ 1 - \frac{8\omega^{2}}{\kappa^{2}} \left( 2 g'^{2} - 1    \right)}  \right]^{1/2},
\end{equation}
and 
\begin{equation}
\Im(\Omega_{\pm}) = \pm \frac{1}{\sqrt{2}}\left[\omega^{2} - \frac{\kappa^{2}}{4} + \frac{\kappa^{2}}{4} \sqrt{ 1 - \frac{8\omega^{2}}{\kappa^{2}} \left( 2 g'^{2} - 1    \right)}   \right]^{1/2},
\end{equation}
respectively. In the large-$\kappa$ limit, we can approximate $\Omega_{\pm}$ to
\begin{equation}
i\Omega_{\pm} \approx \frac{\kappa}{2} \pm i \omega \sqrt{1 - g'^{2}},
\end{equation}
reducing the upper and lower branches of the polariton modes to 
\begin{equation}
\varepsilon^{\mathrm{U}}_{\pm} = \pm i \omega \sqrt{1 - g'^{2}}
\end{equation}
and
\begin{equation}
\varepsilon^{\mathrm{L}}_{\pm} = - \kappa \pm i \omega \sqrt{1 - g'^{2}},
\end{equation}
respectively. In terms of the system's dynamics, the disappearance of the real component of $\varepsilon_{\pm}^{\mathrm{U}}$ allows for a sustained oscillation of $x$ and $y$ in the long-time limit despite the large value of $\kappa$. This closedlike feature of the open LOM in the large-$\kappa$ limit provides us with an alternative picture of how the driven open NOM approaches an effectively closed state in the bad cavity limit without invoking any adiabatic approximation of the cavity mode.

\section{Derivation of LMG Model Semiclassical Equations of Motion} \label{sec:LMG_eom}

In the context of optical cavity-QED experiments as presented in Ref.~\cite{ParkinsPRA}, the master equations of the $\gamma = -1, 0,$ and $1$ LMG model can be written as
\begin{subequations}\label{eq:mastereq}
\begin{equation}
    \partial_t \hat{\rho} = -i[\hat{H}, \rho] + \frac{\alpha^2 \Gamma_a}{N} D[\hat{J}_+] \rho + \frac{\beta^2 \Gamma_b}{N} D[\hat{J}_-] \rho,
\end{equation}
\begin{equation}
    \partial_t \hat{\rho} = -i[\hat{H}, \rho] + \frac{\Gamma_a}{2N} D[2\hat{J}_x] \rho + \frac{ \Gamma_b}{2N} D[\hat{J}_+] \rho,
\end{equation}
and
\begin{equation}
    \partial_t \hat{\rho} = -i[\hat{H}, \rho] + \frac{\Gamma_a}{2N} D[\hat{J}_-] \rho + \frac{ \Gamma_b}{2N} D[\hat{J}_+] \rho,
\end{equation}
\end{subequations}
respectively, where $D[\hat{A}] \hat{\rho} = 2\hat{A} \hat{\rho} \hat{A}^{\dagger} - \{ \hat{A}^{\dagger} \hat{A}, \hat{\rho} \}$, and $\Gamma_k$ are collective atomic dissipation rates related to the cavity field $k$. Specifically, $D[\hat{J}_+] \hat{\rho}$ ($D[\hat{J}_+] \hat{\rho}$) is the dissipation mode related to spontaneous absorption (emission), equivalent to a spin-down (spin-up) state transitioning to a spin-up (spin-down) state. In addition, $D[2\hat{J}_x] \hat{\rho}$ is the dephasing channel, which has negligible effects in the thermodynamic limit.

Instead of working with density operators, we transform Eq.~\eqref{eq:mastereq} into the Heisenberg picture via a transformation similar to Eq.~\eqref{eq:expectvalue_mastereq_DM}, where we use $H_{\text{LMG}}/\hbar$ instead of the LOM Hamiltonian and the dissipation channels are due to the spin raising and lowering operators. By letting $\hat{O} \in \{ \hat{J}_x, \hat{J}_y, \hat{J}_z \}$, we are able to observe how the average values of the collective spin components evolve in time. These, along with the mean-field approximation $\langle \hat{A} \hat{B} \rangle \approx \langle \hat{A} \rangle \langle \hat{B} \rangle$ and rescaling
\begin{equation}
    X = \frac{2\langle \hat{J}_x \rangle}{N}, \quad Y = \frac{2\langle \hat{J}_y \rangle}{N}, \quad, Z = \frac{2\langle \hat{J}_z \rangle}{N},
\end{equation}
allow us to obtain the semiclassical equations of motion of the LMG model expressed in Eq.~\eqref{eq:LMG_eom}, where in particular
\begin{equation}\label{eq:Gammaa}
    \Gamma = \begin{cases}
        \beta^2\Gamma_b - \alpha^2 \Gamma_a \quad &\text{for } \gamma = -1 \\
        -\dfrac{\Gamma_b}{2} &\text{for } \gamma = 0 \\
        \dfrac{\Gamma_a - \Gamma_b}{2} &\text{for } \gamma = 1
    \end{cases}
\end{equation}
recovers the equations of motion based on those from Refs.~\cite{ParkinsPRA, Zhou}. Note that $\Gamma_i < 0$ by definition in Ref.~\cite{ParkinsPRA}; therefore, $\Gamma > 0$, which implies that the dissipation channel for our open LMG model is due spontaneous decay of spins.

\section{Non-DTC Dynamics of the Driven LMG Model}\label{sec:nondtc_lmg}

\begin{figure}[t!]
    \centering
    \includegraphics[width=0.44\textwidth]{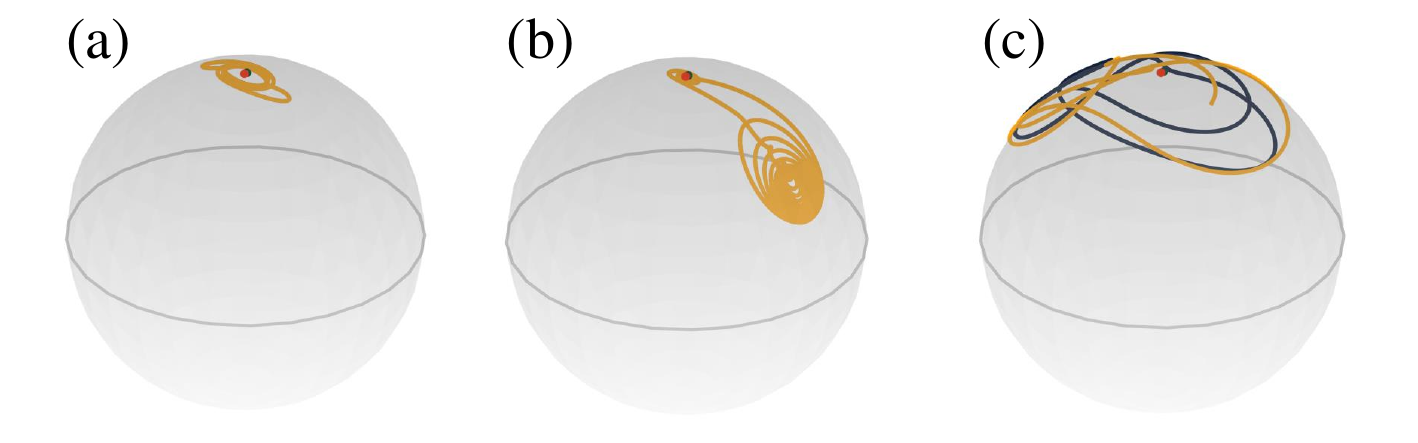}
    \caption{Bloch sphere representation of common non-DTC dynamics in the driven LMG model, i.e., (a) normal beating, (b) symmetry-broken beating, and (c) chaotic behavior, sampled at parameter regions $\{ \lambda_0/\omega_0 , A, 2\omega_d/\omega_0 \} = \{ 0.8, 0.35, 0.5 \}$, $\{ 0.9, 0.9, 0.05 \}$, and $\{ 1.1, 0.6, 1.1 \}$, respectively. The blue (yellow) trajectories mark the dynamics due to an initial state marked by the green (red) dots. All results are for the $\gamma = 0$ and $\Gamma/\omega_0 = 0.1$ case.}
    \label{fig:supp_fig2}
\end{figure}

In this appendix, we discuss additional dynamical non-DTC phases observed in the driven LMG model. Figures~\ref{fig:supp_fig2}(a) and \ref{fig:supp_fig2}(b) present the Bloch sphere dynamics of the normal beating (NB) and symmetry-broken beating (SBB) phases, respectively, analogously named after the NP and SB phase of the undriven model. The collective spins of these phases oscillate about $Z = -1$ and $Z \neq -1$ and have evenly spaced peaks in the frequency spectrum of $X(t)$. They are mostly found alternately appearing in the phase diagrams at resonance lobes with $n > 1$ and in the presence of dissipation, where NB phases appear for odd $n$ and SBB phases appear for even $n$. Some NB phase dynamics at $n>1$ possess higher-order subharmonics with respect to $\omega_d$, but have a time-varying oscillation amplitude, hence the absence of HO-DTCs in the LMG model.

Figure~\ref{fig:supp_fig2}(c) shows the dynamics of a chaotic phase, defined by the sensitivity of the system's behavior to a minute change in its initial conditions and similarly determined as in Sec.~\ref{sec:open_dm} using Eq.~\eqref{eq:decorrelator}. Chaotic phases possess a continuous frequency spectrum, in contrast to other dynamical phases observed in the driven LMG model. They dominate all resonance lobes in the phase diagram in the absence of dissipation.

\bibliography{biblio}

\end{document}